\documentclass[prb,reprint,superscriptaddress]{revtex4-1}
\usepackage[utf8]{inputenc}
\usepackage[T1]{fontenc}
\usepackage{amsmath}
\usepackage{graphicx}
\usepackage{dcolumn}
\usepackage{bm}
\usepackage{epsfig}
\usepackage{float} 
\usepackage{color}
\usepackage{tikz}
\usepackage{subcaption}
\usepackage{makecell}
\usepackage{comment}
\usepackage{cancel}
\usepackage{amssymb}
\usepackage[normalem]{ulem}
\usepackage{multirow}
\newcommand \beq{\begin{eqnarray}}
\newcommand \eeq{\end{eqnarray}}
\newcommand \bea{\begin{eqnarray}}
\newcommand \eea{\end{eqnarray}}

\newcommand \kvec{{\bf k}}

\newcommand \qvec{{\bf q}}

\newcommand\rvec{{\bf r}}

\newcommand\Gvec{{\bf G}}

\def\simge{\mathrel{%
       \rlap{\raise 0.511ex \hbox{$>$}}{\lower 0.511ex \hbox{$\sim$}}}}
\def\simle{\mathrel{
       \rlap{\raise 0.511ex \hbox{$<$}}{\lower 0.511ex \hbox{$\sim$}}}}
\def\beq {\begin{equation}}
\def\eeq {\end{equation}}
\def\w {\omega}
\def\bfq {\mathbf{q}}

\def\bfk {\mathbf{k}}
\def\bfr {\mathbf{r}}

\newcommand{\bra}[1]{\langle #1|}
\newcommand{\ket}[1]{|#1\rangle}
\newcommand{\vo}{V$_2$O$_5$}
\newcommand{\expval}[1]{\langle #1 \rangle}

\newcommand{\rv}{\mathbf{r}}
\newcommand{\kv}{\mathbf{k}}
\newcommand{\qv}{\mathbf{q}}

\newcommand{\gv}{\mathbf{G}}
\newcommand{\rhot}{\tilde{\rho}}

\begin{document}
\title{Robustness of electronic screening effects in electron spectroscopies: example of V$_2$O$_5$}
\author{Vitaly Gorelov}
\affiliation{LSI, CNRS, CEA/DRF/IRAMIS, \'Ecole Polytechnique, Institut Polytechnique de Paris, F-91120 Palaiseau, France}
\affiliation{European Theoretical Spectroscopy Facility}
\author{Lucia Reining}
\affiliation{LSI, CNRS, CEA/DRF/IRAMIS, \'Ecole Polytechnique, Institut Polytechnique de Paris, F-91120 Palaiseau, France}
\affiliation{European Theoretical Spectroscopy Facility}
\author{Walter R. L. Lambrecht}
\affiliation{Department of Physics, Case Western Reserve University, Cleveland,
  OH-441-6-7079, USA}
\author{Matteo Gatti} 
\affiliation{LSI, CNRS, CEA/DRF/IRAMIS, \'Ecole Polytechnique, Institut Polytechnique de Paris, F-91120 Palaiseau, France}
\affiliation{European Theoretical Spectroscopy Facility}
\affiliation{Synchrotron SOLEIL, L’Orme des Merisiers Saint-Aubin, BP 48 F-91192 Gif-sur-Yvette, France}

\date{\today}

\begin{abstract}
In bulk and low-dimensional extended systems, the screening of excitations by the electron cloud is a key feature governing spectroscopic properties. Widely used computational approaches, especially in the framework of many-body perturbation theory, such as the GW approximation and the resulting approximate Bethe-Salpeter equation, are explicitly formulated in terms of the screened Coulomb interaction. In the present work we explore the effect of screening in absorption and electron energy loss spectroscopy, concentrating on the effect of local distortions on the screening and elucidating the resulting changes in the various spectra. Using the layered bulk oxide V$_2$O$_5$ as prototype material, we show in which way local distortions affect the screening, and in which way changes in the screening impact electron energy loss and absorption  spectra including excitons. We highlight cancellations that make many-body effects in the spectra very robust with respect to structural modifications, while the band structure undergoes significant changes and the nature of the excitations may also be affected. This yields insight concerning the structure-properties relations that are crucial for the use of V$_2$O$_5$ as energy storage material, and more generally, that may be used to optimize the analysis and the calculation of electronic spectra in complex materials.
\end{abstract}


\maketitle

\section{Introduction}

Electronic excitations determine materials properties and functionalities that are crucial for a wide range of technological applications. 
Their theoretical analysis and prediction  can therefore accelerate the discovery and design of materials with tailored properties. 
However, this remains a challenging problem for two main reasons: the effect of the electronic interactions and the inherent complexity of real materials. 
On the one hand, the electron-electron interaction strongly affects the excitation spectra of materials  with respect to an independent-particle picture \cite{Martin2016}. 
The Coulomb interaction is, in particular, at the origin of the collective excitations of the electronic charge, known as plasmons \cite{Pines1963}, and it is responsible for the formation of bound electron-hole pairs, termed excitons \cite{Knox1963}. 
Plasmons and excitons characterise the dielectric and optical properties of materials and play a key role as energy carriers for all energy conversion and storage technologies. 
They are the central objects of plasmonics and excitonics, respectively, where their study is of fundamental importance in order to improve the efficiency of optoelectronic devices while reducing their size \cite{Maier2001,Grosso2009}.
On the other hand, real materials differ in several ways from the ideal picture of perfectly periodic crystals that is usually assumed in the theoretical simulation of electronic excitation spectra. 
Besides the thermal motion of the atoms around their equilibrium positions,  or the presence of defects such as vacancies and impurities, static atomic displacements and tiltings can make the atomic local environment different from the macroscopically averaged 
crystal structure that is determined from X-ray diffraction (XRD) data.  
Especially in transition metal compounds, where the electrons remain tightly localised around the ions, the electronic properties  can be strongly affected by different local environments due to positional or magnetic symmetry breakings \cite{Zhao2021,Wang2021,Wang2020,Varignon2019}. 

In this context, vanadium pentoxide {\vo} is a particularly interesting case. 
It is an attractive material for a wide range of environmental applications, such as decontamination treatment, gas sensing, supply of clean and renewable energy (including photocatalysis and photovoltaics),
energy storage (for rechargeable lithium-ion  batteries), and smart windows (thanks to its electrochromic properties)
\cite{Qin2014,Dhayal2010,Razek2020,Monfort2021,Liu2017,Liu2018,Wang2006,Weckhuysen2003,Chernova2009,Lu2017,Fujita1985}.

\begin{figure*}[th]
\center
\begin{minipage}[b]{2.\columnwidth}
\center
       \begin{minipage}[b]{0.6\columnwidth}
        \includegraphics[width=\columnwidth]{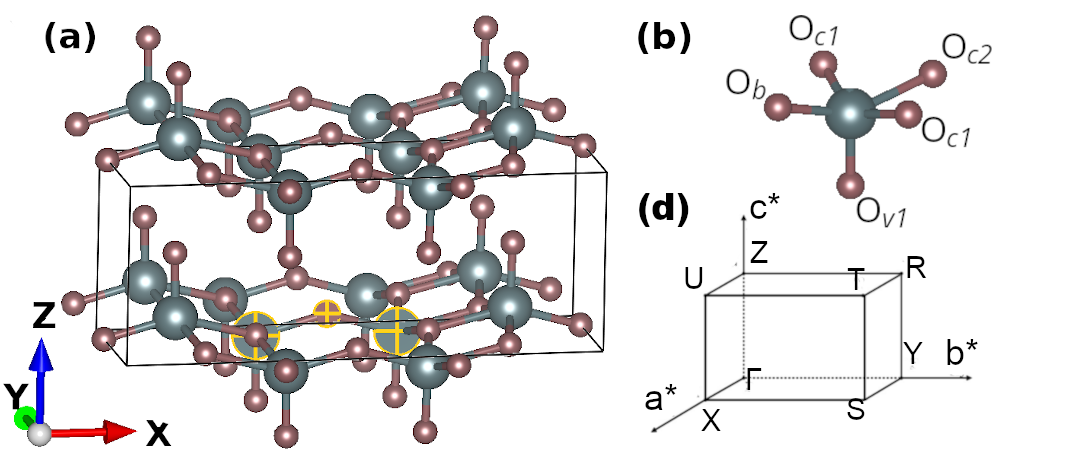}
        \end{minipage}
       \begin{minipage}[b]{0.35\columnwidth}
        \includegraphics[width=\columnwidth]{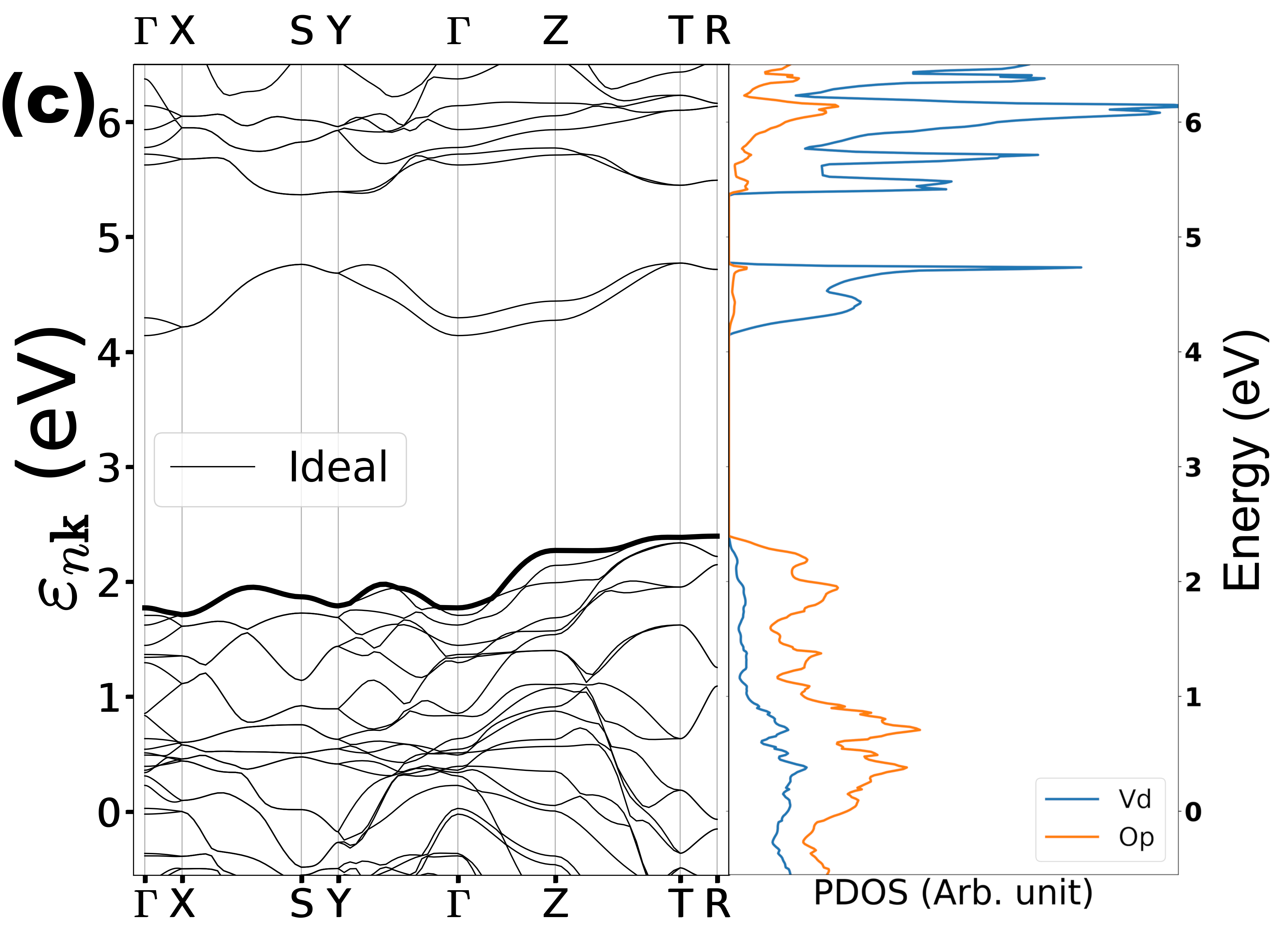}
        \end{minipage}
\end{minipage}
\caption{\small{Crystal structure and electronic properties of {\vo}. (a) {\vo} layers are stacked along the $z$ axis.  V atoms are grey and O atoms are red. Highlighted in yellow is one V-O$_b$-V rung. The orthorhombic unit cell (space group $Pmmn$) is represented by the black lines.
(b) The elementary units of the {\vo} layers are VO$_5$ pyramids, formed by one V atom, two chain oxygen atoms  (O$_{c1}$ and  O$_{c2}$), one bridge oxygen (O$_{b}$) and one vanadyl oxygen (O$_{v1}$). The corresponding bond lengths are reported in Tab. \protect\ref{tab:BONDS}. (c) LDA band structure and corresponding projected density of states (PDOS). The bold line indicates the top valence band (see also Tab. \ref{tab:GWcorr}). (d) The first Brillouin zone with the labels of high symmetry $\bfk$ points. } 
\label{fig:phonon_dist}}
\end{figure*}

{\vo} has a peculiar layered crystal structure \cite{Enjalbert1986}.
The layers, stacked along the $z$ direction, are made of ladder structures with V-O zigzag chains forming the legs along the $y$ direction which are connected by V-O-V rungs along $x$ (see Fig. \ref{fig:phonon_dist}).   
Besides the chain O$_c$ atoms in the legs  and bridge O$_b$ atoms in the rungs, a third kind of oxygen atoms, called vanadyl oxygen O$_v$, are located just above or below the V atoms. 
The resulting basic units are VO$_5$ pyramids formed by a vanadium atom and its five nearest oxygen neighbors (see Fig. \ref{fig:phonon_dist}). 
In each layer, pairs of VO$_5$ pyramids point along the $+z$ direction, alternating with pairs of pyramids pointing in the opposite $-z$ direction.
VO$_6$ octahedra, which are more common in other vanadates, can be formed, although strongly distorted, only by adding to the group also another O$_v$ atom from a neighboring pyramid.

The peculiar crystal structure of {\vo}  is directly reflected into its electronic properties. Its low-dimensional nature is revealed by the weakly dispersing top-valence and bottom-conduction bands, which are separated by a relatively large indirect gap [see Fig. \ref{fig:phonon_dist}(c)]. 
Consistently with the ionic picture of a $d^0$ configuration, the valence band is mainly occupied by O $2p$ electrons, while the conduction band is mainly due to V $3d$ states [as shown by the projected density of states (PDOS) in Fig. \ref{fig:phonon_dist}(c)].
However, O $2p$ and V $3d$ orbitals are also strongly hybridized.
The V-O atomic distances determine the size of the bonding-antibonding splitting, which is larger for the short V-O$_b$ and V-O$_v$ bonds, and smaller for the longer V-O$_c$ bonds (see Tab. \ref{tab:BONDS}). 
The antibonding interactions also depend on the V $3d$ orbital symmetry. As analyzed in Refs. \cite{Lambrecht1981,Bhandari2015},  the V $d_{xy}$ orbitals have the weakest interaction with the nearest oxygen. In particular, the $d_{xy}$ orbitals on the two vanadium atoms across a V-O$_b$-V rung 
that have equal parity are odd with respect to
the $m_x$ mirror plane passing through the bridge O$_b$. Hence, this $d_{xy}$ pair has no interaction with the O$_b$ $p_y$ orbital and is orthogonal to both O$_b$ $p_x$ and O$_b$ $p_z$ orbitals.  This structural peculiarity gives rise to a split-off band separated from the  remaining conduction bands.

A strong interplay between atomic displacements and electronic properties is therefore expected in {\vo} \cite{Smirnov2018}. 
Indeed, Eyert and H\"ock  \cite{Eyert1998} have shown that a hypothetical {\vo} crystal structure built from regular VO$_6$ octahedra, instead of the strongly distorted ones of the real structure, yields a metallic band structure. 
This strong sensitivity has also important practical consequences for lithium-ion batteries. While the processes of
Li intercalation and delithiation do not modify {\vo} lattice parameters substantially, they perturb the local structure of the neighboring {\vo} pyramids, causing considerable band structure changes and a degradation of the electrochemical
performance \cite{Jarry2020,Suthirakun2018,Mukherjee2017,Olszewski2018,Horrocks2016,Delmas1994}. 
Moreover, charge carriers in {\vo} are interpreted as small polarons \cite{Ioffe1970,Scanlon2008,Jesus2016,Watthaisong2019,Ngamwongwan2021}, i.e. the combination of electrons and accompanying lattice distortions, which hinders electronic and ionic mobilities.
Addressing these issues is critical to enhance the performance of {\vo} as a clean energy functional material.

The following question, therefore,  assumes special importance for the physics of {\vo}:
If the atoms of {\vo} are slightly displaced away from the equilibrium positions of the crystal structure determined from XRD data, what is the effect on the electronic properties and the excitation spectra? And are all properties affected in the same manner? If not, are there pieces of information that one could transfer from the ideal material to more complex structures, and can we understand why?
In the present work, we contrast two different kinds of spectroscopy: optical absorption and electron energy loss spectroscopy (EELS), which are popular experimental techniques to determine, respectively, the excitonic and plasmonic properties of materials. 
Addressing this question at the specific example of {\vo}  holds a more general relevance, concerning the impact of structural distortions on dielectric screening and  many-body effects due to the electron-electron and electron-hole interactions.

The present paper is organised as follows.
Sec. \ref{sec:method} summarizes  the methodology and the computational parameters that  have been employed for this study.
The excitation spectra of {\vo} in its ideal crystal structure are first analysed in Sec. \ref{spectra_ideal}.
The impact of  prototypical atomic displacements 
on the absorption spectrum, the dark excitons, and the loss function is then investigated  in Sec. \ref{sec:distortion}.
Finally, Sec. \ref{discussion} discusses the role of screening in various places and the importance of canceling of different contributions, while conclusions are drawn in Sec. \ref{sec:conclusion}.

\section{Methodology} 
\label{sec:method}

\subsection{Theoretical background}

Electronic excitations spectra 
\cite{Onida2002} of {\vo} have been  calculated using  linear-response time-dependent density functional theory   \cite{Runge1984,Ullrich2012}  (TDDFT) and the Bethe-Salpeter equation   \cite{Strinati1988} (BSE) in the framework of many-body perturbation theory  \cite{Martin2016} (MBPT). 

Within TDDFT, the density-density response function $\chi$ 
is obtained as a solution of the Dyson-like equation: 
\beq
\chi = \chi_0 + \chi_0 ( v_c + f_{\rm xc} ) \chi, 
\label{tddft}
\eeq
where $\chi_0$ is the independent-particle response function built with Kohn-Sham (KS) orbitals and energies, $v_c$ is the Coulomb interaction, and 
$f_{\rm xc}$ is exchange-correlation (xc) kernel that is the functional derivative of the KS xc potential $V_{\rm xc}$ with respect to the density. 
The simplest approximation is the random-phase approximation (RPA) that sets  $f_{\rm xc}=0$.  

In reciprocal space, the response functions and the xc kernel  are matrices in the reciprocal lattice vectors $\gv$ and $\gv^\prime$, and functions of the frequency $\w$  and the first-Brillouin-zone wavevector $\qv$. The Coulomb interaction is $v_c(\qv+\gv)=4\pi|\qv+\gv|^{-2}$. 
From the response function $\chi$ one can directly evaluate the inverse microscopic dielectric function:
\begin{equation}
    \epsilon^{-1}_{\gv,\gv'}(\qv,\omega) = \delta_{\gv,\gv'} +
    v_{c}(\qv+\gv)\chi_{\gv,\gv'}(\qv,\omega),
\end{equation}
and the   macroscopic dielectric function from the averaging procedure \cite{Adler1962,Wiser1963}: 
\begin{equation}
\epsilon_M(\qv,\omega) = \frac{1}{\epsilon^{-1}_{\gv=\gv'=0}(\qv,\omega)}.
\label{epsm}
\end{equation}
Eq. \eqref{epsm} takes into account crystal local field effects (LFEs), because the dielectric matrix is inverted before the macroscopic average $\gv=\gv'=0$ is taken. Neglecting LFEs would  simply lead to: $\epsilon_M(\qv,\omega) \simeq \epsilon_{\gv=\gv'=0}(\qv,\omega)$.
The optical absorption spectrum is given by the long-wavelength limit:  $\epsilon_2(\omega) = \lim_{\qv\to 0} {\rm Im}\, \epsilon_M(\qv,\omega)$, whereas the loss function measured by 
EELS as a function of the momentum transfer $\qv$ is given by $-{\rm Im}\, \epsilon_M^{-1}(\qv,\omega)$.
By expressing  the loss function in terms of the real and imaginary parts of the dielectric function $\epsilon_M(\qv,\omega) = \epsilon_1(\qv,\w) + i \epsilon_2(\qv,\w)$, one has: $$-{\rm Im}\, \epsilon_M^{-1}(\qv,\omega) = \frac{\epsilon_2(\qv,\w)}{[\epsilon_1(\qv,\w)]^2+[\epsilon_2(\qv,\w)]^2}.$$
Peaks of the loss function which match zeros of the real part of the dielectric function $\epsilon_1(\qv,\w)$ 
correspond to plasmon resonances.

Within MBPT, the quasiparticle (QP) addition and removal energies that 
form the band structures of materials can be obtained from the poles of the one-particle Green's function $G(\bfr,\bfr',\w)$. 
The effects of the electron-electron interaction beyond the electrostatic Hartree potential are encoded in the 
self-energy $\Sigma_{\rm xc}(\bfr,\bfr',\w)$.
 In the GW approximation \cite{Hedin1965} (GWA), $\Sigma_{\rm xc}(\w)$ 
 is given by the  convolution between $G(\w)$ and the screened Coulomb interaction
$W(\w)=\epsilon^{-1}(\w)v_{c}$ evaluated in the RPA.
In the $G_0W_0$ scheme \cite{Hybertsen1986,Godby1988}, the KS Green's function is used to build the RPA $W_0$ and the self-energy. Moreover, usually the QP energies $E_{n\bfk}$ are obtained as first-order perturbative corrections to the KS eigenvalues $\varepsilon_{n\bfk}$ as:
\beq
E_{n\bfk} = \varepsilon_{n\bfk} + Z_{n\bfk} \bra{\varphi_{n\bfk}}
\Sigma_{\rm xc}(\varepsilon_{n\bfk})-V_{\rm xc}\ket{\varphi_{n\bfk}},
\label{eq:GW}
\eeq
where $\varphi_{n\bfk}$ are KS orbitals and the QP renormalization factors are
$Z_{n\kv}=\left[1 - \expval{\left.\frac{\partial \Sigma_{\rm xc}(\w)}{\partial\w}\right|_{\varepsilon_{n\kv}}} \right]^{-1}$.
Alternatively, QP energies and orbitals entering the GW self-energy are calculated self-consistently, notably within the quasiparticle self-consistent GW (QSGW) scheme\cite{vanSchilfgaarde2006}. While the QSGW results do not depend on the KS starting point, the calculations are computationally much more expensive than the G$_0$W$_0$ ones.

In principle, one can obtain also response properties exactly in MBPT, by solving the Bethe-Salpeter equation.   Approximations are available 
that yield exciton properties of materials accurately \cite{Albrecht1998,Benedict1998,Rohlfing2000}.  In particular, in the GWA with a statically screened Coulomb interaction $W$, the BSE can be reformulated as an electron-hole (excitonic) Hamiltonian problem: $H_{\rm exc} A_\lambda = E_\lambda A_\lambda$.
In the basis $\ket{vc{\bf k}}$ of resonant transitions between occupied  $v{\bf k}$ and  unoccupied  states $c{\bf k}$, and neglecting the coupling with antiresonant transitions  (Tamm-Dancoff approximation),
the excitonic Hamiltonian reads: 
\beq \bra{vc{\bf k}} H_{\rm exc} \ket{v'c'{\bf k}'}=E_{vc{\bf k}} \delta_{vv'}\delta_{cc'}\delta_{{\bf k}{\bf k}'} + \bra{vc{\bf k}} \bar{v}_c-W \ket{v'c'{\bf k}'}.\label{eq:BSE} \eeq
Here $E_{vc{\bf k}}=E_{c\bfk}-E_{v\bfk}$ are the GWA transition energies between occupied and  empty states, while  the electron-hole interaction matrix elements are the sum of the attractive direct electron-hole interaction $-W$ and the repulsive electron-hole exchange interaction $\bar v_c$  given by the microscopic components (i.e., ${\bf G}\neq 0$) of the bare Coulomb interaction. 
In the Tamm-Dancoff approximation, 
which is usually a good approximation for semiconductors and insulators \cite{Onida2002},
the optical absorption spectrum  is obtained from the eigenvectors $A_\lambda^{vc{\bf k}}$ and eigenvalues $E_\lambda$ of the excitonic hamiltonian as: 
\beq
\epsilon_2(\w) =  \lim_{\qv\to 0}\frac{8\pi^2}{\Omega q^2} \sum_\lambda \left|\sum_{vc{\bf k}}  A_\lambda^{vc{\bf k}} \tilde{\rho}_{vc{\bf k}}(\qv) \right|^2 \delta(\w- E_\lambda ),
\label{spectrumBSE2}
\eeq
where $\Omega$ is the crystal volume, and the oscillator strengths are
$\rhot_{vc{\bf k}}(\qv)= \int \varphi^*_{v\kv-\qv}(\rv) e^{-i\qv\cdot\rv}\varphi_{c\kv}(\rv) d\rv$.
With respect to the independent-particle approximation where the electron-hole interactions are neglected, peaks in the BSE absorption  spectra are located at the exciton energies $E_\lambda$, instead of $E_{vc{\bf k}}$, and have modified intensities resulting from the mixing of the  oscillator strengths that are modulated by the excitonic coefficients   $A_\lambda^{vc{\bf k}}$.
In this context, the exciton binding energy is defined as the energy difference between the minimum direct gap energy $E_{vc{\bf k}}$ and the exciton energy $E_\lambda$.
Supposing that in the $\bfq\to0$ limit 
the oscillator strengths $\sum_{vc{\bf k}}  A_\lambda^{vc{\bf k}} \tilde{\rho}_{vc{\bf k}}$ are approximately proportional to $1/E_{\lambda}$, Eq. \ref{spectrumBSE2} reduces to $\epsilon_2(\w) \propto \textrm{JDOS}(\w)$, with the joint density of states $\textrm{JDOS}(\w)=\frac{2}{\Omega \w^2} \sum_\lambda \delta(\w- E_\lambda)$.

\subsection{Computational details}

In our calculations, the ideal crystal structure of {\vo}, which is shown in Fig. \ref{fig:phonon_dist},  has a $Pmmn$ orthorhombic symmetry  with 14 atoms per unit cell, and the experimental lattice parameters \cite{Enjalbert1986} $a=11.512$ \AA{}, $b=3.564$ \AA{},  and $c=4.368$ \AA{}.  
This reference crystal structure will be compared to less symmetric unit cells, where the atomic positions have been displaced, as detailed in Sec. \ref{sec:distortion}.
In all simulations we have used norm-conserving Troullier-Martins \cite{Troullier1991}  pseudopotentials, including  $3s$ and $3p$ semicore states for vanadium in the valence (total of 112 electrons), which have been already validated in previous studies on vanadates \cite{Papalazarou2009,Gatti2015}. 
The KS ground-state calculation, in the local density approximation (LDA), converged with an energy cutoff of 100 Hartree and a $4\times4\times4$ $\bfk$-point grid.

TDDFT calculations were performed within the RPA using 145 conduction and 41 valence bands.  When comparing to the loss function calculated with the computationally more expensive BSE (Fig. \ref{fig:EELS_analys}) we have used $4\times4\times4$ $\bfk$-point grid; for the comparison of the calculated loss function to experiment (Fig. \ref{fig:EELS}) we have used a $6\times6\times6$ $\bfk$-point grid, and a $8\times8\times8$ $\bfk$-point grid to see the fine differences between the  loss functions of ideal and distorted structures (Fig. \ref{fig:EEL_dist}).
The $\bfk$-point convergence study is reported in the Supplementary Material\cite{suppmat} (SM). 

In the perturbative GW calculations, the dielectric matrix within a cutoff energy of 4.9 Hartree was computed within the RPA with LDA energies that were scissor-corrected  to match the G$_0$W$_0$ band gap  at the $\Gamma$ point, using a $4\times4\times4$ $\bfk$-point grid and 350 bands, while the self-energy required 700 bands and 52 Hartree cutoff energy.

The BSE spectra have been obtained with a $4\times4\times4$ $\bfk$-point grid, with 15 valence and 16 conduction bands for absorption spectra (resulting in the converged spectra up to 7 eV), and 41 valence and 60 conduction bands for EELS spectra that were converged up to 23 eV.
The BSE hamiltonian,  Eq. \eqref{eq:BSE}, has been built, following the standard approach, using results from the perturbative  GW calculations 
(with LDA wave functions and scissor corrected LDA energies)
and a statically screened interaction $W$ calculated in RPA with scissor-corrected LDA energies. 
In Ref. \cite{Gorelov2022} the BSE hamiltonian for the ideal {\vo} crystal structure was built on the basis of an expensive QSGW calculation.
The computational scheme employed here slightly reduces the exciton binding energy (e.g. from 1.0 eV to 0.8 eV in the case of the bright exciton, see first row of Tab. \ref{tab:exc-dist}). However, most importantly, it gives the same physical picture of excitonic effects and makes the study at the same time meaningful and computationally feasible for the several distorted crystal structures analysed in the present work.
More in general, all computational ingredients have been checked in this sense, where we could benefit from our results in Ref.\cite{Gorelov2022}, which contains a detailed analysis of the sensitivity of the absorption spectra with respect to the ingredients and computational choices in the BSE. 
We used this to set the frame which allows us in the present work to make comparisons that are significant, but obtained at a reasonable computational cost. App. \ref{app:kpoint} and App. \ref{app:scissor} discuss in detail, respectively,  the $\bfk$-point convergence and the validity of the scissor correction approximation.
A 0.15 eV Gaussian broadening was applied to the energy loss spectra and 0.1 eV to the absorption spectra.
LDA and GW calculations were carried out with ABINIT \cite{Gonze2005}, while the DP \cite{DPcode} and EXC codes \cite{EXCcode} were used for  TDDFT and BSE calculations, respectively.

\section{Results}

\subsection{Spectra of the ideal crystal}
\label{spectra_ideal}

\subsubsection{Absorption spectra and exciton properties}

Absorption spectra of the ideal {\vo} crystal structure have been recently investigated in Ref.  \cite{Gorelov2022} by us. Here we summarize the main findings and we refer to  Ref.  \cite{Gorelov2022} for a more extended discussion.

Our BSE calculations in Ref.  \cite{Gorelov2022} showed that the first peak at the onset of the absorption spectra is a tightly bound exciton, which is characterised by a large average electron-hole distance. 
The properties of this exciton are therefore in contrast with the textbook picture of the Frenkel exciton model, where a large exciton binding  energy is associated with a strong localisation of the electron-hole pairs\cite{Knox1963,Bassani1975}. 
We explained this unusual behavior by means of a tight-binding model representing the charge transfer nature of the exciton that stems from electron-hole transitions within the  V-O$_b$-V rungs having a mirror symmetry.  
The solution of the model gives an exciton wavefunction localised in the $y$ and $z$ directions but delocalised along $x$. Within the model, in this direction its extension is much larger than the V-O$_b$-V distance and does not depend on the strength of the electron-hole interaction.

The BSE calculation also  revealed the presence of a pair of dark excitons (one for each V-O$_b$-V rung in the unit cell) below the onset of the optical spectrum. 
Even though they have a negligible dipole oscillator strength and therefore are not visible in the absorption spectrum, dark excitons play an important role in disexcitation processes like as phonon-assisted luminescence and light emission \cite{Feierabend2017,Park2017,Kusaba2021}.
In {\vo} the dark excitons result from the same charge transfer transitions within the V-O$_b$-V rungs  as the bright excitons at the optical onset and are described by the same tight binding model. 
These excitons are dark because of the destructive interference of single-particle transitions that are instead individually bright. Our tight-binding model showed that this is a consequence of the peculiar crystal symmetry of {\vo}. Therefore, of particular interest here is what happens to the dark excitons when the structure is distorted and the symmetry broken.

\subsubsection{Electron energy loss spectra}

\begin{figure*}[t]
\center
\begin{minipage}[b]{2.\columnwidth}
\includegraphics[width=0.8\columnwidth]{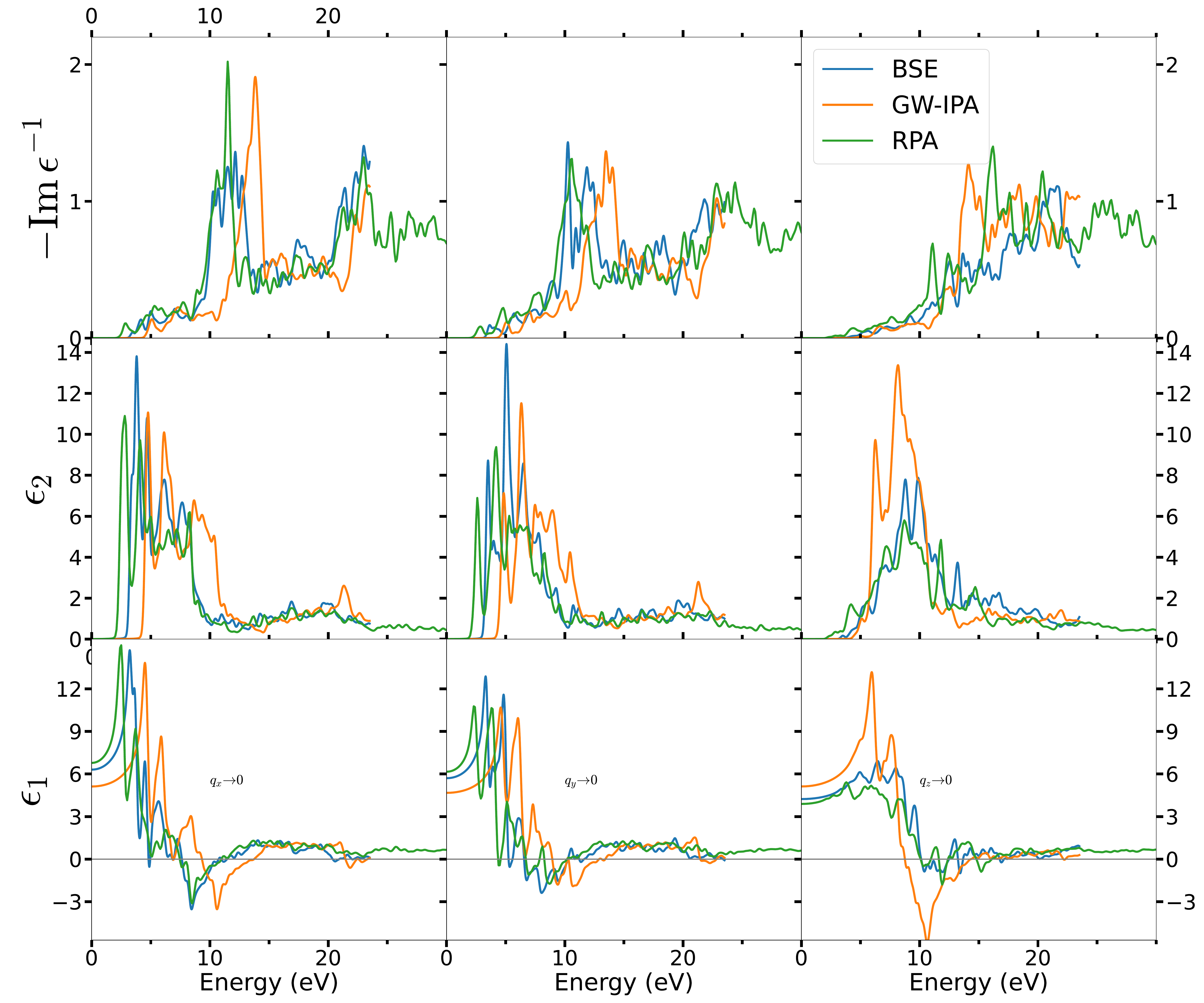}
\end{minipage}
\caption{\small{Loss function  $-\text{Im} \epsilon^{-1}_M(\bfq\to0,\omega)$ (top panels) and corresponding imaginary part $\epsilon_2(\bfq\to0,\w)$ (middle panels), and real part $\epsilon_1(\bfq\to0,\w)$ (bottom panels) of the macroscopic dielectric function
of {\vo}
in the 3 cartesian directions $x$ (left), $y$ (center) and $z$ (right),  computed within the TDDFT-RPA (green), the GW-IPA (orange) and the BSE (blue). 
}  \label{fig:EELS_analys}}
\end{figure*}

While Ref. \cite{Gorelov2022} dealt with the exciton properties and the absorption spectrum of {\vo}, 
here we also investigate its loss function. Its analysis provides a significant complementary viewpoint on the impact of local atomic distortions on many-body effects.
Indeed, the loss function is dominated by plasmon excitations, which are usually associated with  the long-range nature of the Coulomb interaction. 
Moreover, EELS allows for the study of the wavevector dependence of the electronic excitations, well beyond the dipole limit probed by optical spectroscopy.

The top panels of Fig. \ref{fig:EELS_analys} compare the loss functions in the long-wavelength limit
$-\textrm{Im} \,\epsilon^{-1}_M(\qv\rightarrow0,\w)$
calculated from the solution of the BSE
with the spectra obtained within the GW independent-particle approximation (IPA)
and the RPA of TDDFT.
While in the latter the independent-particle energies in $\chi_0$, see Eq. \eqref{tddft}, are KS eigenvalues in the LDA, in the GW-IPA they are corrected by a rigid scissor that opens the band gap, simulating the GW corrections. The GW-IPA, however, completely neglects electron-hole interactions.

\begin{figure*}[t]
\center
\begin{minipage}[b]{2.\columnwidth}
\includegraphics[width=\columnwidth]{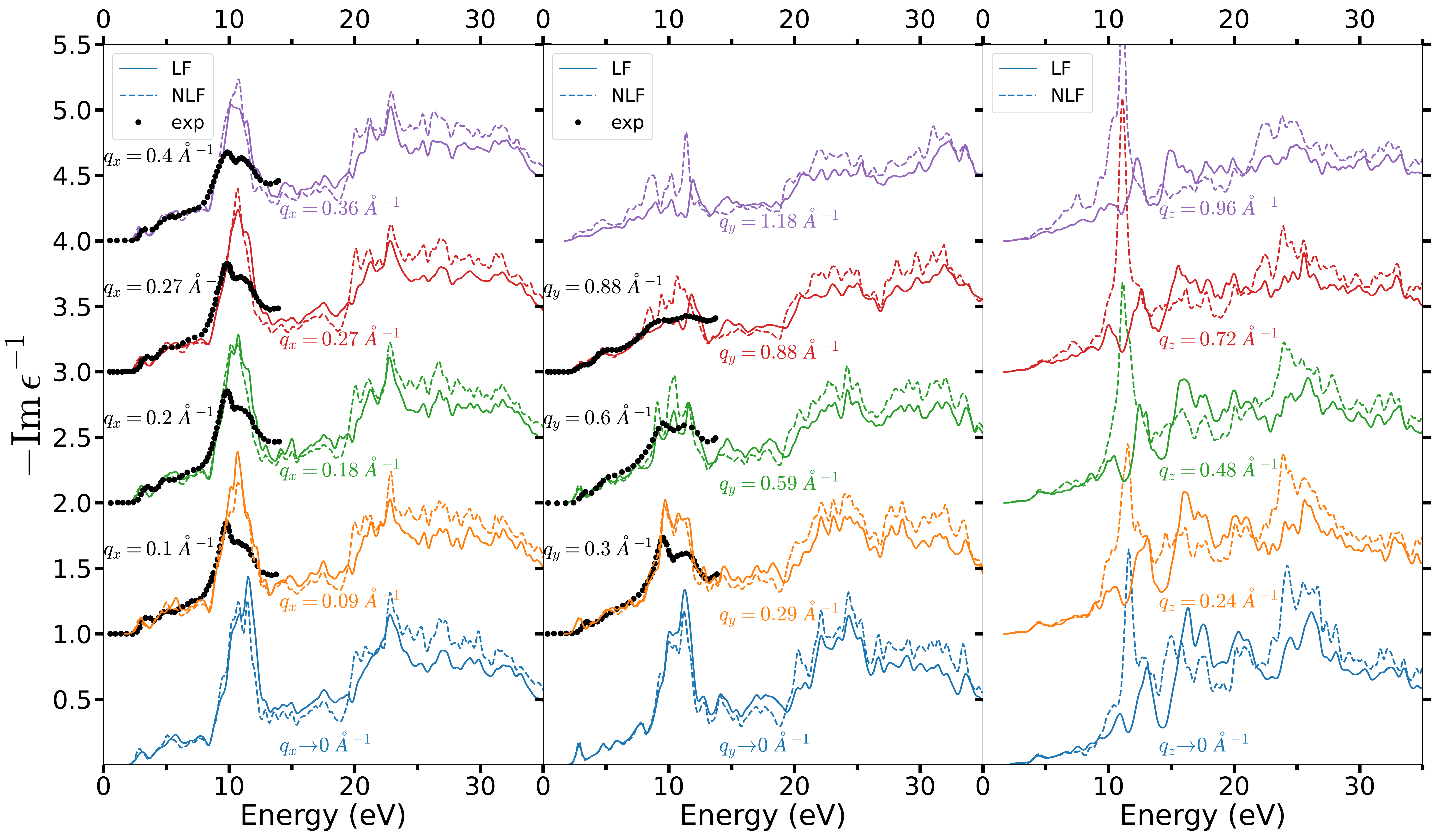}
\end{minipage}
\caption{\small{Dispersion of the loss function $-\text{Im}\epsilon_M^{-1}(\bfq,\w)$ as a function of momentum transfer $\bfq$ in the 3 cartesian directions, computed within TDDFT-RPA with (solid lines) and without (dashed lines) crystal local field effects, and comparison to EELS experimental data \cite{Atzkern2000} (black dots).  A vertical
offset has been added to the spectra for improved clarity. The size of the theoretical or experimental momentum transfer is reported next to each spectrum. 
}
\label{fig:EELS}}
\end{figure*}

 In comparison to the absorption spectra, excitonic effects are less prominent in the loss function. 
In the $x$ and $y$ directions, the GW-IPA spectra  are mainly a blueshift of the RPA spectra, as an effect of the GW band-gap opening. 
Electron-hole interactions in the BSE have two main effects: they make the main peak at 11 eV  broader, and they induce a redshift that partially compensates the difference between RPA and GW-IPA at the onset of the spectra.
As a result of these cancellations, RPA spectra are much more similar to BSE than GW-IPA spectra.
Most importantly, excitonic effects do not create new strong peaks within the band gap, since the dominant excitonic peaks in the absorption spectra are strongly renormalized, and become insignificant, in the EELS. 
In the $z$ direction, besides those global shifts, a redistribution of spectral weight is also apparent between GW-IPA spectra on one side and RPA  and BSE spectra on the other side. This is a manifestation of crystal local fields that will be analysed more in detail in the following.
In any case, here we can already conclude that the main features in the loss function in {\vo} are adequately described at the level of the RPA, which will therefore also be adopted  to investigate the dispersion of the loss function  as a function of the momentum transfer.

From the analysis of  the RPA and BSE loss functions in terms of the real and imaginary parts of the dielectric function shown in the middle and bottom panels of Fig. \ref{fig:EELS_analys}, we find that the 11 eV peak in the loss functions  corresponds to a zero of  $\epsilon_1(\qv\to0,\w)$. Therefore it can be ascribed to a collective plasmon resonance rather than to an interband electron-hole excitation \cite{Atzkern2000}.  
The spectral features at lower energies in the loss function are instead associated to peaks of $\epsilon_2(\qv\to0,\w)$ and can therefore be
considered to stem from interband transitions between the top-valence   O $2p$ bands and the V split-off conduction bands.
Finally, the very broad peak at 
larger energies around 23 eV  matches a minimum of $\epsilon_1(\qv\to0,\w)$, which however never crosses the zero axis. We can therefore assign it to a plasmon that is strongly damped by the interaction with single-particle excitations.
While the plasmon at 11 eV is due to excitations of O $2p$ electrons into the V $3d$ empty bands, the  broad plasmon at 23 eV is due to excitations to higher energy empty bands.

\begin{figure*}[t]
\center
\begin{minipage}[b]{2.\columnwidth}
\center
        \begin{minipage}[b]{0.7\columnwidth}
        \includegraphics[width=\columnwidth]{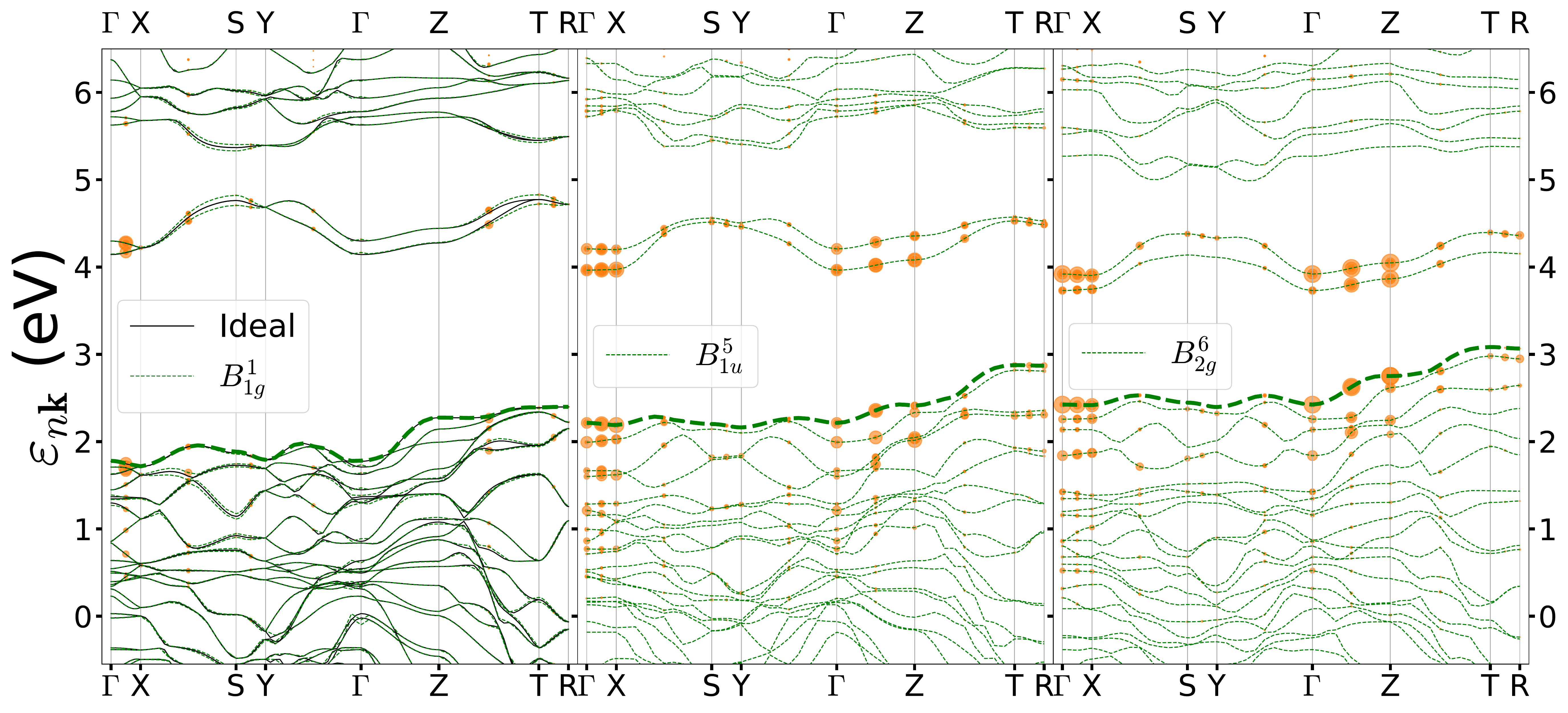}
        \end{minipage}
\end{minipage}
\caption{\small{LDA band structures of the ideal and distorted structures. The bold dashed lines indicate the top-valence band. The considered displacement corresponds to the largest one  in Tab. \protect\ref{tab:BONDS}. The band structure of the crystal distorted along the $B^1_{1g}$ mode (left-most panel) almost entirely overlaps that of the ideal structure [see Fig. \protect\ref{fig:phonon_dist}(c)].
The size of the orange circles is proportional to the contribution $|A_\lambda^{vc\kv}\tilde\rho_{vc\kv}|$ to the lowest energy exciton in the $y$ polarization direction (see text).} 
\label{fig:BS_dist}}
\end{figure*}

Fig. \ref{fig:EELS} shows the RPA loss function $-\textrm{Im} \epsilon^{-1}_M(\qv,\w)$ for several momentum transfers $\qv$ along the 3 cartesian directions,  calculated  with or without the inclusion of local field  effects (LFEs) over an extended energy range. 
The calculations are compared to the experimental EELS spectra of Ref. \onlinecite{Atzkern2000}.
For $x$ and $y$ directions, LFEs affect the spectra especially at large energies and have a stronger impact for increasing size of the momentum transfer. 
In the $z$ direction, which is perpendicular to the {\vo} layers, the inhomogeneities are larger. Therefore, LFEs are much stronger, inducing essentially a blueshift of the spectra.
As a result, the spectra along $z$ are much more affected by LFE than the spectra along the in-plane $x$ and $y$ directions where the inhomogeneity is smaller. 
For the  momentum transfers and in the whole energy range for which experimental results are available, the agreement with the calculated spectra is very good. 
Both the simulations and the experiment in the $x$ and $y$ directions display a main peak at $\sim$ 11 eV.
The main peak is preceded by smaller spectral features and an onset located at $\sim$ 3 eV.   
All the peaks show little dispersion as a function of momentum transfer reflecting the localised character of the electronic excitations.
The intensity of the main peak at $\sim$ 11 eV turns out to be systematically larger in the calculations than in the experiment.
We note that that the momentum resolution in the EELS experiment\cite{Atzkern2000} is  as large as 0.05-0.06 {\AA}$^{-1}$. Therefore, the resulting intensity of the measured peaks may have been strongly reduced in the experiment as a consequence of the angular integration. 
As a matter of fact, it is plausible that the experimental spectra for $\bfq$ along $x$ and $y$ in reality also contain contributions coming from the out-of-plane $z$ direction, where at the same energy loss the calculated loss function does not display any intense feature. 
This also calls for further measurements with improved angular resolution, also for momentum transfers $\bfq$ along $z$, where experimental data are presently not available.

\subsection{Effects of distortions}
\label{sec:distortion}

To study the impact of atomic displacements in {\vo} in detail, in the following we compare the spectra calculated in the ideal crystal structure with those obtained by slightly displacing the atoms of {\vo} inside the unit cell. The displacements we have chosen were inspired by related observations, in particular, phonon modes.
However, it is important to stress that we do not claim that our distorted structures would correspond to crystals that are closer to realistic samples than the ideal structure, not least because we keep the unit cell unchanged. Instead, they may be considered to be prototype distortions that allow us to analyse representative effects on absorption and loss function, both in view of the effects of distortions that are observed in applications based on nanostructured V$_2$O$_5$, and to draw general conclusions about the impact of local distortions in different kinds of spectroscopy.

\subsubsection{Away from the ideal crystal}

We consider deformations of the crystal structure of {\vo} from the experimental X-ray diffraction (XRD) positions \cite{Enjalbert1986} along the phonon modes calculated in Ref. \onlinecite{Bhandari2014}. In the following, we show spectra obtained in the structures distorted along three representative vibrational modes: $B^1_{1g}$, $B^5_{1u}$ and  $B^6_{2g}$.

While the $B^1_{1g}$ mode contains only displacements along the $y$ direction,  notably for V and O${}_v$ atoms, both the $B^5_{1u}$ and $B^6_{2g}$ modes contain $x$ and $z$ motions. In particular, the $B^5_{1u}$ mode mainly corresponds to a stretch of the V-O$_c$ bond, while $B^6_{2g}$ is a V-O$_b$ bond-stretch mode with a motion of the O$_b$ atoms opposite to each other \cite{Bhandari2014}. The phonon frequencies, $\omega_v$, of these modes are: 147 cm$^{-1}$ for $B^1_{1g}$, 963 cm$^{-1}$ for $B^6_{2g}$ (from Raman spectroscopy data)\cite{Clauws1985} and 570 cm$^{-1}$ for $B^5_{1u}$ (from infrared spectroscopy \cite{Clauws1976}). While these phonons except for $B_{1g}^1$ are not contributing significantly to the thermal motion, they may still have an effect via the zero-point-motion electron-phonon  coupling band gap and exciton renormalization.  The characteristic length of a phonon mode  $v$, $l_v =\sqrt{\hbar/2M \omega_v}$  is about 0.08 {\AA} for the $B_{2g}^6$ mode and 0.11 {\AA} for the $B_{1u}^5$ mode.  (In the chosen modes, the maximum displacement  corresponds to the O atom, so we used the O mass in these order of magnitude estimates.) These are comparable in magnitude to the arbitrarily chosen amplitudes here.  However, beside possible zero-point motion corrections, within this work we are primarily concerned with the static local distortions, which may arise for various reasons, including sample preparation or nanostructuration. Therefore, within a reasonable range, the amplitudes of displacements are chosen arbitrarily conserving the relative motion of atoms along the corresponding phonon modes. 
For the $B^5_{1u}$ and  $B^6_{2g}$ modes we have also considered a small and a large distortion (for each structure the largest displacements $d_{\max}$ are indicated in the first column of Tab. \ref{tab:BONDS}). 
For the distortions that we have examined in the present work, the bond length changes are reported in Tab. \ref{tab:BONDS}.
All the atomic positions of the deformed structures are listed in the SM \cite{suppmat}. The visualization of phonon modes can be found in Ref.\cite{Bhandari2014}.

\begin{table}[ht]
    \centering
    \begin{tabular}{c|c|c|c|c}
    \hline\hline
         Structure & V-O$_b$ & V-O$_{c2}$ & V-O$_{c1}$ & V-O$_{v1}$ \\ 
          \hline
        Ideal  & 1.78 & 2.02 & 1.88 & 1.58 \\ 
                 \hline
         $B^1_{1g}$ $d_{max}=0.03$ \AA & 1.78 & 2.02 & 1.86 1.89 & 1.58 \\ 
                   \hline
        $B^5_{1u}$  $d_{max}=0.02$ \AA& 1.77 1.79 & 1.99 2.05 & 1.88      & 1.58 \\ 
        $B^5_{1u}$  $d_{max}=0.12$ \AA& 1.72 1.84 & 1.85 2.19 & 1.87 1.89 & 1.58 \\ 
                     \hline
        $B^6_{2g}$ $d_{max}=0.04$ \AA & 1.72 1.83 & 2.02 & 1.88 & 1.58 \\ 
        $B^6_{2g}$ $d_{max}=0.13$ \AA & 1.63 1.92 & 2.00 2.04 & 1.88 & 1.57 1.59 \\ 
        \hline\hline
    \end{tabular}
    \caption{\label{tab:BONDS} \small{Bond lengths of the ideal and distorted {\vo} structures (in \AA). 
    In the ideal structure, for each kind of bond, the length values are all the same. In the distorted structures, instead, they are not all equivalent anymore: the shortest and longest atomic distances for each kind of bond are therefore reported. }
    }
\end{table}

\subsubsection{Effects of distortions on the band structure}

\begin{figure*}[ht]
\center
\begin{minipage}[b]{2.\columnwidth}
\center
        \begin{minipage}[b]{\columnwidth}
        \includegraphics[width=0.9\columnwidth]{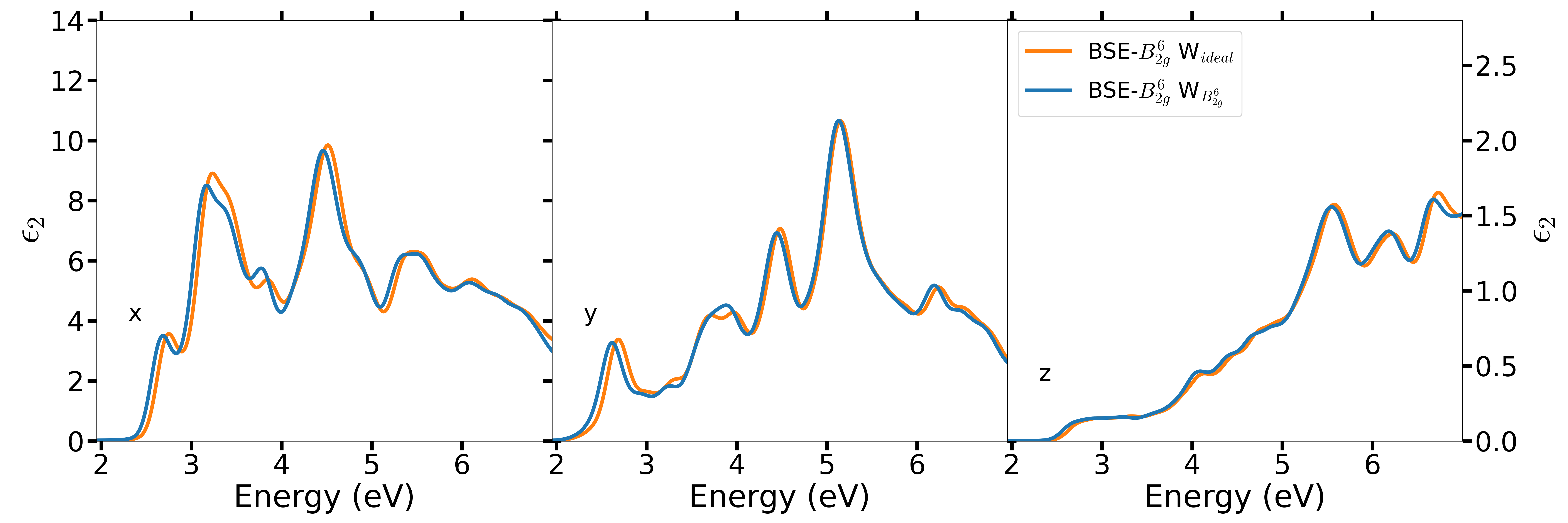}
        \end{minipage}
\end{minipage}
\caption{ The absorption spectra, in the 3 cartesian polarization directions, (left) $x$, (middle) $y$ and (right) $z$, of the crystal structure distorted along the $B_{2g}^6$ mode, for the case of the largest displacements using the BSE with $W_{\rm ideal}$ (orange lines) or the screened interaction $W_{\rm dist}$ of the distorted structure (blue lines). 
\label{fig:ABS_B2g6}}
\end{figure*}

Fig. \ref{fig:BS_dist} shows the LDA band structures  
of the ideal and distorted structures  in the regime of the largest atomic displacements. The LDA band gap\footnote{We note that  the LDA gives similar results for {\vo} to the generalised gradient approximation  \cite{Perdew1996} (GGA) (see also e.g.  Refs. \cite{Wang2006b,Jovanovic2018,Das2019}). In particular, for the ideal structure, we obtain a GGA band gap of 1.84 eV.} of the ideal structure is 1.8 eV.
In the $B^1_{1g}$ case  (left-most panel) the band structure almost entirely overlaps that of the ideal {\vo} crystal.
In all the other cases, instead, 
the band structure is strongly affected, with the band dispersions that are largely modified everywhere.
As a consequence the LDA indirect band gap is reduced and goes from 1.74 eV for the $B^1_{1g}$ 
to 1 eV for $B^5_{1u}$ and  0.65 eV for $B^6_{2g}$ case. 
The changes are the largest for the $B^6_{2g}$ case (third panel in Fig. \ref{fig:BS_dist}), where the two narrow bottom conduction bands are split everywhere across the Brillouin zone. 
The bands around the band gap therefore turn out to be very sensitive to the 
V-O$_b$ bond stretch, distinctive of the 
$B^6_{2g}$ phonon mode, and, to a smaller extent, to the 
V-O$_c$ bond stretch, associated to 
$B^5_{1u}$ phonon mode.
They are instead quite stable with respect to displacements along $y$ of the O$_v$ atoms ($B^1_{1g}$ mode).
In the next section, we will examine how these modifications of the band structures due to the crystal deformations are reflected into changes of the optical properties.

\subsubsection{Absorption spectra of the distorted crystal}
\label{sec:abs}

\begin{figure*}[ht]
\center
\begin{minipage}[b]{2.\columnwidth}
\center
        \begin{minipage}[b]{\columnwidth}
        \includegraphics[width=0.8\columnwidth]{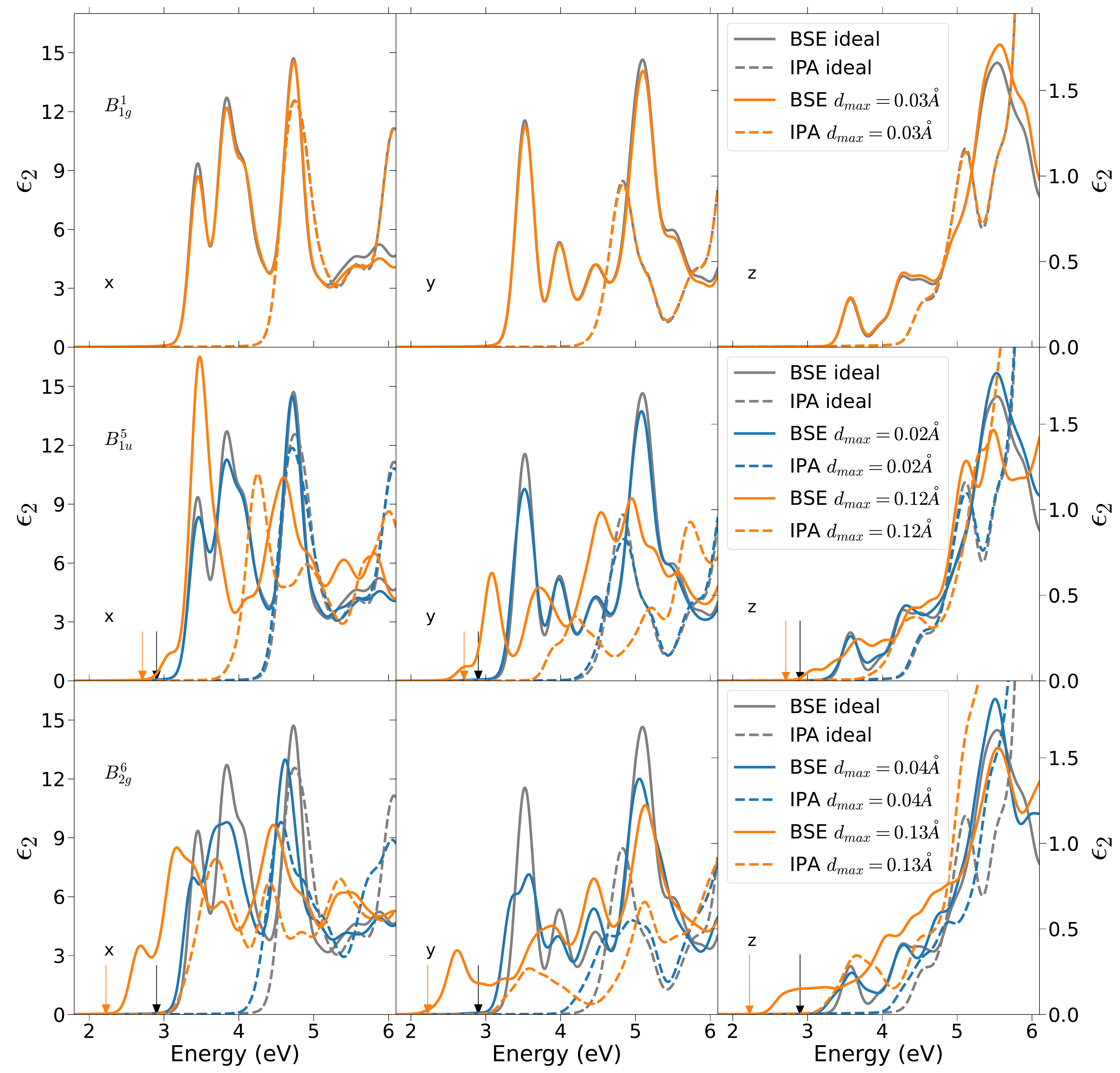}
        \end{minipage}
\end{minipage}
\caption{Absorption spectra of the distorted crystal structures (orange and blue lines) compared to the absorption spectra in the {\vo} ideal structure (gray lines). The spectra are obtained from the solution of the
BSE (solid lines) and from the GW-IPA (dashed lines), where e-h interactions are neglected.
The $d_{max}$ values are the largest atomic displacement in the unit cell with respect to the ideal crystal (see Tab. \protect\ref{tab:BONDS}). For the $B^5_{1u}$ and $B^6_{2g}$ phonon modes both a small (blue lines) and large (orange lines) displacement are considered.
The arrows for the structures distorted along the $B_{2g}^6$ and $B_{1u}^5$ modes indicate the lowest exciton energy, which is perfectly dark in the ideal {\vo}  structure (black arrows), but not in the $y$ direction in the distorted structures (orange arrows). 
}
\label{fig:ABSdist}
\end{figure*}


For both GW and BSE calculations, an expensive part of the computation is the evaluation of the screened Coulomb interaction $W$.
In principle, its calculation should be repeated for each of the different distorted structures.
In Fig. \ref{fig:ABS_B2g6} we analyse a possible shortcut for the BSE absorption spectra of the structure with the largest distortion for the $B_{2g}^6$ mode.
It is interesting  to examine here this specific case more in detail since its  band structure is the most affected by the crystal deformation.

In particular, we assess the possibility to employ the screened Coulomb interaction  $W_{\rm ideal}$ of the ideal structure instead of $W_{\rm dist}$ of the distorted structure.
The spectra remain very similar 
(compare blue and orange spectra): using $W_{\rm ideal}$ at the place of $W_{\rm dist}$ does not alter significantly the absorption spectra of the $B_{2g}^6$  distorted structure.
On the basis of these observations, we can conclude that we can safely calculate the absorption spectra of all the other distorted crystal structures (where the changes in the band structures are  smaller than for the $B^6_{2g}$ case) using
the screened interaction $W_{\rm ideal}$ of the ideal crystal structure.
We will analyse and explain these observations in detail in Sec. \ref{discussion}.

Fig. \ref{fig:ABSdist} compares the absorption spectra $\epsilon_2(\w)$,  calculated along the 3 cartesian polarization directions, for all  distorted crystal structures to the spectra of the ideal {\vo}. 
To examine how the excitonic properties of {\vo} change when the crystal structure is modified, we first consider the situations where the distortions along the phonon modes are small, with the largest displacements between $d_{\rm max} \sim 0.02$ {\AA} and  $d_{\rm max} \sim 0.04$ \AA.    
While the $B^1_{1g}$ mode has a negligible effect on the spectra, for the $B^5_{1u}$ mode and especially the $B^6_{2g}$ mode the changes are sizable in all directions (compare gray and  blue lines in Fig.  \ref{fig:ABSdist}). 
For the  larger crystal distortions along the $B^5_{1u}$  and $B^6_{2g}$ phonon modes,  there are conspicuous changes in the spectra (compare gray and orange solid lines in Fig.  \ref{fig:ABSdist}): the  energy of the first bright exciton is redshifted by up to 0.7 eV  and the shape of the lowest energy peaks is strongly modified.

\begin{table}[th]
    \centering
   \begin{tabular}{c|c|c|c|c}
    \hline\hline
          Structure &
          Optical  & Direct  & $E_b$ & $E_b$  \\
           & onset & bandgap & bright & dark \\ 
           \hline
       Ideal  & 
        3.4 & 4.2 & 0.8 & 1.3 \\ 
                 \hline
         $B^1_{1g}$ $d_{max}=0.03$ \AA & 
         3.4 &     4.2 & 0.8 & 1.3  \\ 
                   \hline
        $B^5_{1u}$  $d_{max}=0.023$ \AA& 
         3.4 & 4.2 & 0.8 & \\ 
        $B^5_{1u}$  $d_{max}=0.12$ \AA & 
        3.1 & 3.8 & 0.7 & 1.1  \\ 
                     \hline
        $B^6_{2g}$ $d_{max}=0.04$ \AA  & 
        3.4 & 4.1 &   0.7 & \\ 
        $B^6_{2g}$ $d_{max}=0.13$ \AA  & 
        2.7 & 3.3 &  0.6 & 1 \\ 
                  \hline
         \hline\hline
    \end{tabular}
    \caption{\label{tab:GAPS} \small{Energy of the first bright exciton (termed "Optical onset") for $x$ polarization from BSE calculation, smallest direct band gap from scissor-corrected LDA and exciton binding energy (in eV) of the lowest energy bright and dark excitons for the ideal and distorted {\vo} structures. 
    }
    }
\end{table}

By comparing the GW-IPA spectra (dashed lines), we observe that the effects of the distortions are already largely visible: the BSE spectra (solid lines) actually follow the same trends of the GW spectra. 
The redshifts of the energies of the first bright excitons, see Tab. \ref{tab:GAPS}, are the combined effect of two changes:  the shrinkage of the GW direct band gap and the reduction of the exciton binding energy. We find that the band gap change gives the largest contribution to the spectral redshift, while the binding energies of both the lowest energy bright and dark excitons remain comparatively more stable.
We will analyse the origin of these changes more in detail in Sec. \ref{discussion}.

For the $B^5_{1u}$ and $B^6_{2g}$ modes, the BSE spectra (orange solid lines)  in the $y$ direction also develop a tail on the low energy side of the first prominent peak (see orange arrows). 
This points to the fact that the lowest energy exciton that is perfectly dark in the ideal {\vo} structure (see black arrows)
 has some non-vanishing oscillator strength in the $y$ direction in these distorted structures.
 
This lowest energy exciton in the $y$-polarization shifts to 2.3 and to 2.7 eV when applying the $B_{2g}^6$ and $B_{1u}^5$ phonon distortions respectively. The $\epsilon_2$ for these two excitons is about 25 and 13 times smaller than for the first bright exciton in this polarization for the perfect crystal, and this is for phonon amplitudes that are still somewhat larger than our estimate of the corresponding phonon length scale $l_{B_{2g}^6}$ and $l_{B_{1u}^5}$ from zero-point-motion. However, this is still compatible with the 2 orders of magnitude lower absorption at the 2.3 eV onset than for the bright exciton near 3 eV. in Ref. \onlinecite{Kenny1966}. Thus, the often quoted lower band gap extracted from the onset of absorption could plausibly be ascribed to brightening of the dark exciton by zero-point motion electron-phonon renormalization of the exciton.  While this is clearly not a complete calculation of the exciton-phonon coupling effect, it is still a plausible reason for the lower onset of absorption near 2.3 eV. Alternatively, indirect band gap excitons could also contribute again by electron phonon coupling.

\subsubsection{Brightening of the dark excitons}

\begin{figure}[ht]
\center
\includegraphics[width=0.68\columnwidth]{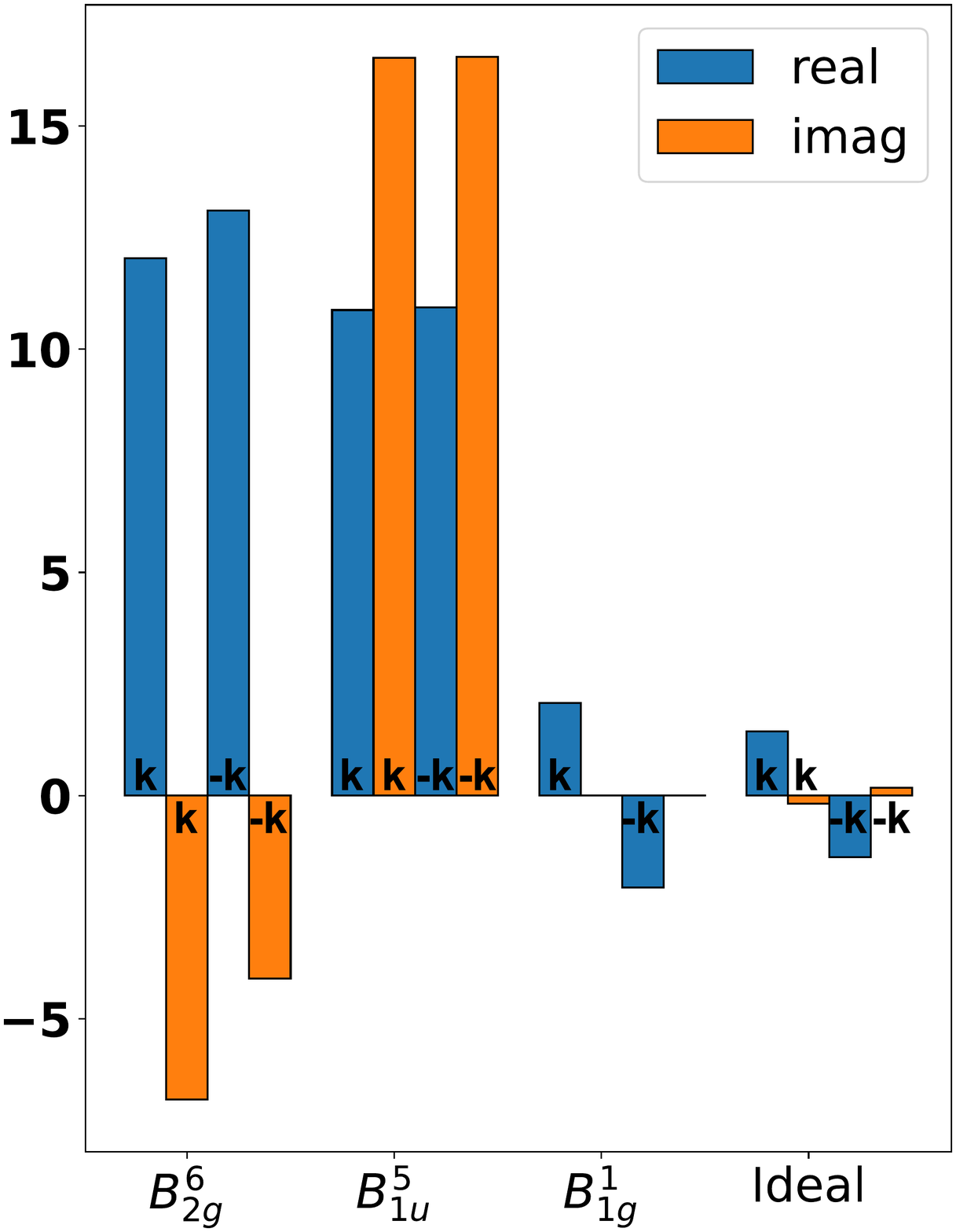}
\caption{\small{Contributions to the oscillator strength of the lowest energy exciton in the $y$ polarization direction for the crystal structures distorted along  $B^6_{2g}$ (first histograms from the left), $B^5_{1u}$ (second histograms), $B^1_{1g}$ (third histograms) modes in comparison to the ideal  {\vo} case (fourth histograms). Blue (orange) columns represent the real (imaginary) parts of $\sum_{vc \kv} A_\lambda^{vc\kv}\tilde\rho_{vc\kv}$, where the sum over $\bfk$ is restricted to one half of the first Brillouin zone. The $\bfk$ points belonging to each subset are related to those of the other subset by inversion symmetry  $\kv\leftrightarrow -\kv$. 
In the $B^1_{1g}$ and ideal structures, the columns have same heights but opposite signs, so the exciton is dark. In the other distorted structures, the columns are higher and their signs do not cancel, so the exciton becomes bright. 
}\label{fig:Analys_dark}}
\end{figure}

\begin{figure*}[ht]
\center
\begin{minipage}[b]{2.\columnwidth}
\center
        \begin{minipage}[b]{\columnwidth}
        \includegraphics[width=0.8\columnwidth]{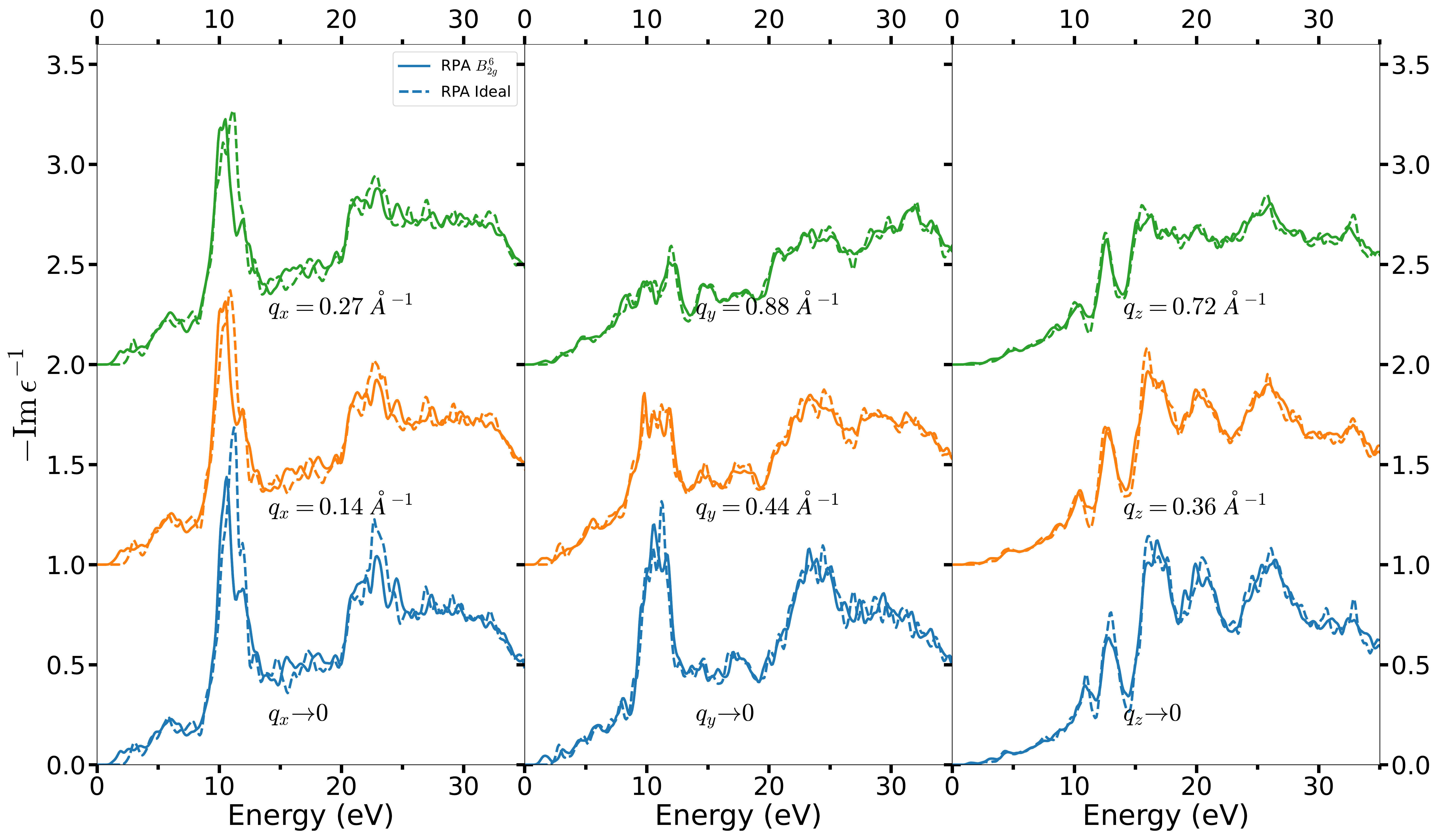}
        \end{minipage}
\end{minipage}
\caption{\small{TDDFT-RPA loss function spectra $-\text{Im}\epsilon_M^{-1}(\bfq,\w)$ as a function of the momentum transfer $\bfq$ along the $x$ (left), $y$ (middle) and $z$ (right) directions,
calculated  for the structure distorted along the $B^6_{2g}$ mode (solid lines) and the undistorted structure (dashed lines).  A vertical
offset has been added to the spectra for improved clarity. The size of the   momentum transfer is reported next to each spectrum 
}\label{fig:EEL_dist}}
\end{figure*}

The oscillator strength of each exciton $\lambda$ is given in Eq. \eqref{spectrumBSE2} by the square modulus of the sum over transitions $v\bfk\to c\bfk$ of the products $A_\lambda^{vc\kv}\tilde\rho_{vc\kv}$.
Therefore, in order to understand how the dark exciton of the ideal {\vo} structure becomes bright for the $y$ polarization direction
in the crystal structures distorted along the  $B^5_{1u}$ and  $B^6_{2g}$ modes, we analyse the contributions  $A_\lambda^{vc \kv}\tilde\rho_{vc\kv}$ to the exciton oscillator strength.
As shown in Ref.\cite{Gorelov2022}, in the ideal crystal structure the dark exciton is the result of the destructive interference of contributions $A_\lambda^{vc\kv}\tilde\rho_{vc\kv}$  stemming from  $\kv$ points that are equivalent by symmetry, each term $A_\lambda^{vc\kv}\tilde\rho_{vc\kv}$ being separately different from zero. 
Fig. \ref{fig:BS_dist} represents the individual contribution $|A_\lambda^{vc\kv}\tilde\rho_{vc\kv}|$ in the band structure plot for the distorted {\vo} structures. 
There we consider specifically the $y$ polarization of matrix elements for the lowest energy exciton.  
Each transition $v\bfk \to c\bfk$ is represented by a pair of orange circles, one in a valence band and the other in a conduction band, at the same $\bfk$ point. The size of the circle is proportional to the weight of each contribution. We find that there are many contributions different from zero, mainly involving the top-valence and bottom-conduction bands.
The $B^1_{1g}$ structure (left-most panel in Fig. \ref{fig:BS_dist}), where the exciton remains dark, is the closest to the ideal structure. In particular, it preserves the inversion symmetry.
The $B^5_{1u}$ and  $B^6_{2g}$  distorted structures (see second and third panels Fig. \ref{fig:BS_dist}), instead, only preserve the 
$m_y$ reflection symmetry  with respect to the $y$ axis.
In these two distorted structures the individual contributions $|A_\lambda^{vc\kv}\tilde\rho_{vc\kv}|$ become larger than in the ideal  case and, in addition, stem from high-symmetry $\bfk$ points (e.g. X, Z, T and $\Gamma$).
Notably, in the ideal and $B^1_{1g}$ structures $|A_\lambda^{vc\kv}\tilde\rho_{vc\kv}|$ is zero at the $\Gamma$ point, which is not equivalent by symmetry to other $\bfk$ points in the first Brillouin zone.
Interestingly, in $x$ direction, instead, the corresponding $|A_\lambda^{vc\kv}\tilde\rho_{vc\kv}|$
in the distorted crystal structure remain  similar to the ideal case and come from the same part of the Brillouin zone (see SM\cite{suppmat}).

To illustrate the cancellation, associated with the inversion symmetry $\bfk \to -\bfk$,  of the individual contributions in the resulting absorption spectrum, we have divided the Brillouin zone in two half volumes, either containing  a $\bfk$ point or its corresponding $-\bfk$. 
We have then restricted the sum over $\bfk$ points in Eq. \eqref{spectrumBSE2} to either half of the Brillouin zone. In the ideal crystal structure, the two semi-sums  are small, but still different from zero (see the rightmost histograms in Fig. \ref{fig:Analys_dark}). However, they are equal and with an opposite sign, which implies that they cancel when summed together, giving rise to the dark exciton with negligible oscillator strength. 
This holds true also for the $B^1_{1g}$ case, whereas in the  $B^6_{2g}$ and $B^5_{1u}$ distorted structures (see two leftmost histograms in Fig. \ref{fig:Analys_dark})  the semi-sums have much larger intensities, and, most importantly, they  have the same sign.
As a result, in these distorted structures the dark exciton becomes visible at the onset of the absorption spectra. In particular, for the $B^5_{1u}$ structure the real and imaginary parts have the largest intensity, which also explains why the corresponding peak has a larger intensity in the spectrum.

\subsubsection{Electron energy loss spectra of the distorted crystal}

Fig. \ref{fig:EEL_dist} shows the RPA loss functions for several momentum transfers along the 3 cartesian directions calculated in the  structure distorted along the $B^6_{2g}$ mode, where the effect on the absorption spectrum is the largest. These loss functions are compared to the corresponding spectra calculated in the ideal {\vo} crystal structure.
For all the momentum transfers, the spectra of the distorted structure remain very similar to the ideal case. In particular, the plasmon resonances at 11 and  23 eV are not strongly affected by the local distortions: we observe a small redshift by 0.5 eV and a minor reduction of intensity of the peaks.
The largest modifications  occur in the low-energy region, for $\w < 10$ eV, where the peaks with lower intensity are associated to spectral features $\epsilon_2$ and therefore undergo similar changes  as in the absorption spectra. 
In other words,
while the absorption spectra are largely modified by the structural distortions, mainly through changes in the underlying band structure, the loss function spectra are much more stable.
We can understand this striking difference between the two excitation spectra by the fact that plasmons are collective excitations of the electronic charge that are  mainly linked to properties of the global environment like the average electron density and are much less sensitive to changes of bond lengths. 
In the next section, we will therefore analyse more in detail the origin of the changes in the absorption spectra.

\section{Discussion}
\label{discussion}

Our main focus of interest here is to analyze the effect of structural distortions on the many-body effects in the spectra.
We consider  in particular the representative case of the structure distorted according to the $B^6_{2g}$ mode 
with the larger atomic displacement of 0.13 {\AA}.

\subsection{Screening in the short- and long-range}

\begin{figure}[t]
\center
\begin{minipage}[b]{\columnwidth}
\center
        \begin{minipage}[b]{\columnwidth}
        \includegraphics[width=0.8\columnwidth]{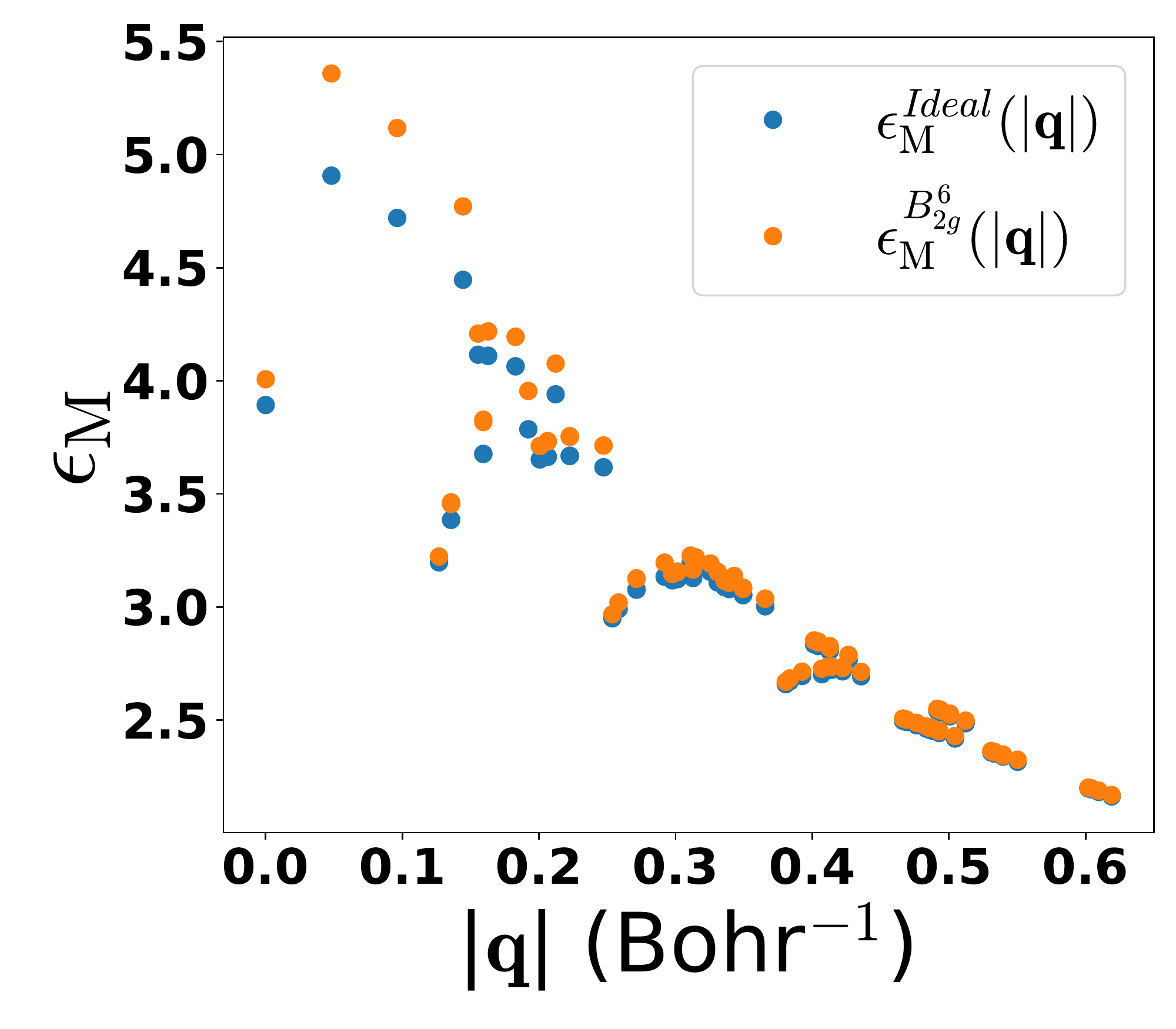}
        \end{minipage}
\end{minipage}
\caption{\small{Static   macroscopic  dielectric function  $\epsilon_M(\qvec,\w=0)=1/\epsilon^{-1}_{\Gvec=0,\Gvec'=0}(\qvec,\w=0)$  for ideal and distorted structures (blue and orange dots, respectively) calculated within the RPA. The values at $|\mathbf{q}| \to 0$ have been calculated along the direction (1,2,3) in reciprocal lattice units and are meant to represent an average of the different spatial directions. 
}
\label{fig:SCR_q}}
\end{figure}

On the level of the GW approximation, the effective interaction entering both the self-energy and the electron-hole interaction is the screened Coulomb interaction $W$. The computation of the screening is the most expensive part of the calculations, and at the same time, the interpretation of changes in screening upon changes in the material is not straightforward. It is therefore worthwhile to examine it more in detail.
Fig. \ref{fig:SCR_q} shows the  static macroscopic dielectric constant $\epsilon_M(\qvec,\w=0)$, defined in Eq. \eqref{epsm}, that screens the Coulomb interaction in momentum space. 
 $\epsilon_M$ takes different values for the same or similar $|\qvec|$; this is due to the inhomogeneity and the anisotropy of {\vo}. Especially the anisotropy explains the peculiar shape of the distribution, since the screening perpendicular to the planes is significantly weaker than the in-plane screening. As one would expect, at large $|\bfq|$, where short distances are probed, any dielectric function tends to 1 and therefore the distorted and undistorted systems give similar values. Moreover, the dependence on the direction of $\bfq$ is weak. At smaller $\bfq$ screening increases, and with this, a stronger direction dependence arises and distorted and undistorted results start to differ. This is particularly clear between 0.2 and 0.05 ${\rm Bohr}^{-1}$. The curve seems to drop at $\bfq\to 0$, but this is merely due to the fact that we show here a particular direction of $\bfq$, namely (1,2,3) in reciprocal lattice units.  The figure suggests that results will be most affected by the changes in screening upon distortion when they are dominated by smaller $\bfq$, hence, by larger distances. It is therefore interesting to analyse the role of distortions in the GW and BSE results.

\subsection{Consequences of the screening}

\subsubsection{In the $GW$ quasiparticle corrections}

The changes in the GW QP corrections 
- see Eq. \eqref{eq:GW} - 
can come through changes of the wavefunctions used in the calculations of matrix elements, changes of Kohn-Sham energies in the one-body Green's function, and the changes of the screening observed above. 
In Tab. \ref{tab:GWcorr} we first compare the main matrix elements at the $\Gamma$ point that result from the GW calculation for the top valence (tv) and bottom conduction (bc) states, for the ideal and distorted structures, see rows \textcircled{1} and \textcircled{2}, respectively. 
In the table, the GW matrix elements are broken up into the different contributions adding up in  Eq. \eqref{eq:GW}, which we analyse separately.

\begin{table*}[tb]
\centering
    \begin{tabular}{l|c|c|c|c|c|c|c|c}
    \hline\hline
        Structure &  Band &  $\varepsilon_{n\bfk}$ & $\langle V_{\rm xc}\rangle$ &  $\langle\Sigma_{\rm x}\rangle$ & $\langle\Sigma_{\rm c}(\varepsilon_{n\bfk})\rangle$  & $Z_{n\bfk}$ &   $E_{n\bfk}-\varepsilon_{n\bfk}$  &  $E_{n\bfk}$ \\
         \hline
       \multirow{2}{*}{\textcircled{1} Ideal}  &    tv  & 1.78 & -19.88& -24.24 &  4.02 &  0.78&     -0.27 &  1.51  \\
 &  bc &  4.14& -22.30& -14.22&  -5.57 &  0.77 &   1.95  & 6.09
 \\
          \hline
        \multirow{2}{*}{\textcircled{2} Distorted (with  $W_{\rm dist}$)} & tv & 2.42 & -19.51& -23.97 & 4.23 &  0.77 &   -0.17 &  2.25 \\
     & bc &  3.73& -22.34& -14.41 & -5.40 &  0.77 &   1.95 &  5.68 \\
        
  \hline
        \multirow{2}{*}{\textcircled{3} Distorted   (with  $W_{\rm ideal}$)} &  tv  & 2.42 & -19.51& -23.97 & 4.16 & 0.77 & -0.22 & 2.20  \\
               &  bc &  3.73 & -22.34& -14.41 & -5.32&   0.77&   2.02 &  5.75
 \\
        \hline\hline
    \end{tabular}
    \caption{\label{tab:GWcorr} \small{Individual contributions to the GW QP corrections  - see Eq. \protect\eqref{eq:GW} -  for the top-valence (tv) and bottom conduction (bc) bands at $\bfk=\Gamma$ for the ideal {\vo} and the crystal structure distorted along the $B^6_{2g}$ mode with largest atomic displacement $d_{\rm max}=0.13$ \AA. For the distorted structure two calculations are compared:  using the screened Coulomb interaction $W$ calculated either in the distorted structure $W_{\rm dist}$ (second row) or in the ideal structure  $W_{\rm ideal}$ (third row). 
    }}
\end{table*}

In the third column, the KS energies $\varepsilon_{n\bfk}$ are given, with the noteworthy distortion-induced changes consistent with the band structure in Fig. \ref{fig:BS_dist}. The bare exchange matrix elements $\langle\Sigma_{\rm x}\rangle$, which depend only on wave functions, are shown in the fifth column. The distortions lead to an upshift of the valence band and a downshift of the conduction band by about 0.2 eV, which represents only 1 and 2\% of $\langle\Sigma_{\rm x}\rangle$ for the valence and conduction state, respectively. The increased exchange of the conduction state points to an on average stronger overlap of conduction  and valence states upon distortion, which may stem from a slight average delocalization of the valence state that would explain its reduced exchange matrix element.
Screening enters the correlation contribution $\langle\Sigma_{\rm c}(\varepsilon_{n\bfk})\rangle$ to the self-energy in the sixth column. This contribution can be split into an overall downwards shift due to the short-range Coulomb hole, and a contribution that reduces the exchange downshift, i.e., an upshift, which is dominant for the valence band where exchange is stronger. The cancellations result in a correlation contribution that exhibits changes of similar magnitude as the bare exchange, in spite of the changed screening. The energy dependence of the self-energy, measured by the quasiparticle renormalization factor $Z_{n\bfk}$ in the seventh  column, is marginally affected. 
The resulting total self-energy matrix element of the conduction state does not reflect any significant changes upon distortion, whereas the valence band self-energy changes by about half an eV. However, the matrix elements of the KS potential $\langle V_{\rm xc}\rangle$ in the forth column show the very same trends, which means that the final quasiparticle correction $E_{n\bfk}-\varepsilon_{n\bfk}$   changes by less than 0.1 eV. It is very important to stress that there is indeed a noteworthy change of the valence band energy due to changes in the xc effects upon distortions, but that this change is already captured by the Kohn-Sham potential. 

The main computational effort in the GW calculation is the evaluation of screening. The row \textcircled{3} of Table \ref{tab:GWcorr} shows the results obtained using the distorted structure, but with $W_{\rm ideal}$ computed in the ideal crystal. 
Being able to use  $W_{\rm ideal}$ throughout would indeed constitute a significant computational gain.
Therefore, we now compare rows \textcircled{2} and \textcircled{3}. In order to evaluate directly the quality of the approximation, one has to check the matrix element of the correlation self-energy $\langle\Sigma_{\rm c}(\varepsilon_{n\bfk})\rangle$  in the sixth column of Table \ref{tab:GWcorr}. The calculation using the $W_{\rm ideal}$ captures about 70\% of the effect of distortion on the valence band, whereas it overestimates the distortion effects by about 50\% on the conduction band. In other words, the distortion effects are captured to a significant extent, though not with high precision. It should be noted, however, that the differences here are of the order of 0.1-0.2 eV, which starts to be in the range of the precision of the GW calculation itself. Moreover, as pointed out above, the final QP correction $E_{n\bfk}-\varepsilon_{n\bfk}$ is very well described even by a fully ideal calculation, thanks to the fact that the effects of distortion is already well captured in the Kohn-Sham calculation.

\subsubsection{In the BSE calculations}

In the BSE calculation of the exciton the e-h interaction has no Kohn-Sham counterpart, and one would therefore not expect cancellations. This would mean that distortions could lead to major changes and require a recalculation of $W$.  
Indeed, the optical gaps and exciton binding energies in Table \ref{tab:GAPS} show that distortions can induce significant changes.  The distortion of 0.13 {\AA} along the $B^6_{2g}$ mode leads to a 0.9 eV decrease of the  direct band gap at the  $\Gamma$ point: it is accompanied by a decrease of the  energy of the first bright exciton of 0.7 eV, which means, the binding energy of the bright exciton decreases by 0.2 eV, from 0.8 to 0.6 eV. The binding energy of the dark exciton shows a similar decrease of 0.3 eV. This is a relatively moderate change of the binding energy, and one might want to conclude quickly that the excitons do not feel the distortion strongly.

However, a closer look incites to be cautious. We first consider the matrix elements of $W$ calculated using the exciton wavefunction in a state $\lambda$:
\bea
    \langle W_{\lambda} \rangle 
    &=& \sum_{\substack{vc\kvec \\ v'c'\kvec'}} A_\lambda^{*vc{\bf k}} \langle vc \kvec | W |v'c'\kvec' \rangle A_\lambda^{v'c'{\bf k'}}\,.
    \label{eq:w-lambda}
\eea
The rows and columns of Tab. \ref{tab:W} identify the possible choices,  respectively, for the 
eigenvectors $A_\lambda^{vc{\bf k}}$ and the matrix elements $\langle vc \kvec | W |v'c'\kvec' \rangle$  that enter Eq. \eqref{eq:w-lambda}.
In the  row \textcircled{1}, the expectation value \eqref{eq:w-lambda}  is calculated using the  eigenvectors $A_\lambda^{vc{\bf k}}$, for the bright and dark excitons,  obtained from the excitonic hamiltonian of the ideal structure.
In the row \textcircled{2}, instead, the excitonic eigenvectors $A_\lambda^{vc{\bf k}}$ are calculated in the distorted structure (where the excitonic hamiltonian contains a screened interaction $W$ that is also calculated in the distorted structure). 
In third column, named $\langle W_\lambda \rangle_{\rm ideal}$, the  matrix elements $\langle vc \kvec | W |v'c'\kvec' \rangle$ of the  screened Coulomb interaction calculated in the ideal structure enter Eq. \eqref{eq:w-lambda}.
In the forth column $\langle W_\lambda \rangle_{\rm dist}$, instead, they are the matrix elements $\langle vc \kvec | W |v'c'\kvec' \rangle$ calculated in the distorted structure.

\begin{table}[t]
    \centering
    \begin{tabular}{l|c|c|c|c|c|c}
    \hline\hline
        Structure  & Exciton  &  $\langle W_\lambda \rangle_{\rm ideal}$ & $\langle W_\lambda \rangle_{\rm dist}$ & $E^{\rm eff}_{\lambda}$ & $E_{\rm g}$ & $E_b$ \\
          \hline
 \multirow{2}{*}{\textcircled{1} Ideal} & Dark  & 1.9  &  &  4.7 & \multirow{2}{*}{4.2} &  1.3 \\
         & Bright  & 1.7  &  & 5.0 &  & 0.8\\
                 \hline
 \multirow{2}{*}{\textcircled{2}          Distorted } & Dark  & 1.6  & 1.6 & 3.7 & \multirow{2}{*}{3.2} & 1  \\
         & Bright  & 1.0  & 1.0 & 3.5 &  & 0.6\\
                 \hline
 \textcircled{3}          Distorted  & Dark  & 1.6  & 1.6 & 3.9 & \multirow{2}{*}{3.3} & 1  \\  
         (with $W_{\rm ideal}$) & Bright  & 1.0  & 1.0 & 3.6 &  & 0.6  \\   
    \hline \hline
    \end{tabular}
    \caption{\label{tab:W}\small{Matrix elements (in eV) of the screened Coulomb interaction, see Eq. \eqref{eq:w-lambda}, in the ideal  and distorted  crystal structures ($\langle W_\lambda \rangle_{\rm ideal}$ and $\langle W_\lambda \rangle_{\rm dist}$, third and forth columns, respectively),  and the effective gap $E^{\rm eff}_{\lambda}$  (fifth column), see Eq. \eqref{eq:eff_gap},, calculated using the exciton wavefunctions  obtained in the ideal and distorted structures (second and third rows, respectively), for the dark and bright excitons. The fifth an sixth columns report the gaps and the exciton binding energies.
    In the bottom  row, the exciton wavefunctions are obtained in the distorted crystal structure using in the BSE hamiltonian the screened interaction $W_{\rm ideal}$ from the ideal structure.}} 
       \label{tab:exc-dist}
\end{table}

We first consider the difference $\langle W_\lambda \rangle_{\rm ideal}$-$\langle W_\lambda \rangle_{\rm dist}$ between rows \textcircled{1}  and \textcircled{2}.
This expectation value difference should give a rough estimate of the change of exciton binding energy, and indeed, in the case of the dark exciton the distortion decreases $\langle W_\lambda \rangle$ from 1.9 to 1.6 eV, i.e., by 0.3 eV, right the decrease from 1.3 to 1.0 eV of the binding energy $E_b$ (see last column of Tab. \ref{tab:W}). 
However, the same is not true for the bright exciton: here $\langle W_\lambda \rangle$ decreases as much as 0.7 eV (from 1.7 to 1.0 eV), which  is almost four times the 0.2 eV decrease of the binding energy. This indicates that there is a significant change in the bright exciton wave function $A_{\lambda}^{vc\bfk}$, although this cannot be detected in its binding energy.  

Tab. \ref{tab:exc-dist} also compares  $\langle W_\lambda \rangle$ from Eq. \eqref{eq:w-lambda}
 with the direct gap $E_g$ (sixth column)
 and the effective gap  $E^{\rm eff}_{\lambda}$ (fifth column), defined as:
\begin{equation}
    E^{\rm eff}_{\lambda} = \sum_{vc\bfk}|A_{\lambda}^{vc\bfk}|^2(E_{c\bfk}-E_{v\bfk})\,,
    \label{eq:eff_gap}
\end{equation}
i.e., the expectation value of the conduction-valence energy difference in the excitonic state of interest. Indeed, neglecting the weak local field effects, the exciton binding energy is given by: 
\beq
E_b=E_g -\left(E^{\rm eff}_{\lambda} -\langle W_{\lambda} \rangle\right),
\label{eq:Eb}
\eeq
rather than by $\langle W_{\lambda} \rangle$ alone. Even for our low-lying excitons, the difference between the minimum direct gap $E_g$ and the effective gap $E^{\rm eff}_{\lambda}$ can be significant: it is 0.8 eV for the bright exciton in the ideal structure, as the  row \textcircled{1} of Table \ref{tab:exc-dist} shows. In order to understand the change in binding energy upon distortion, we therefore also have to examine the change of effective gap. 

\begin{figure*}[ht]
\center
\begin{minipage}[b]{2.\columnwidth}
\center
        \begin{minipage}[b]{\columnwidth}
        \includegraphics[width=0.7\columnwidth]{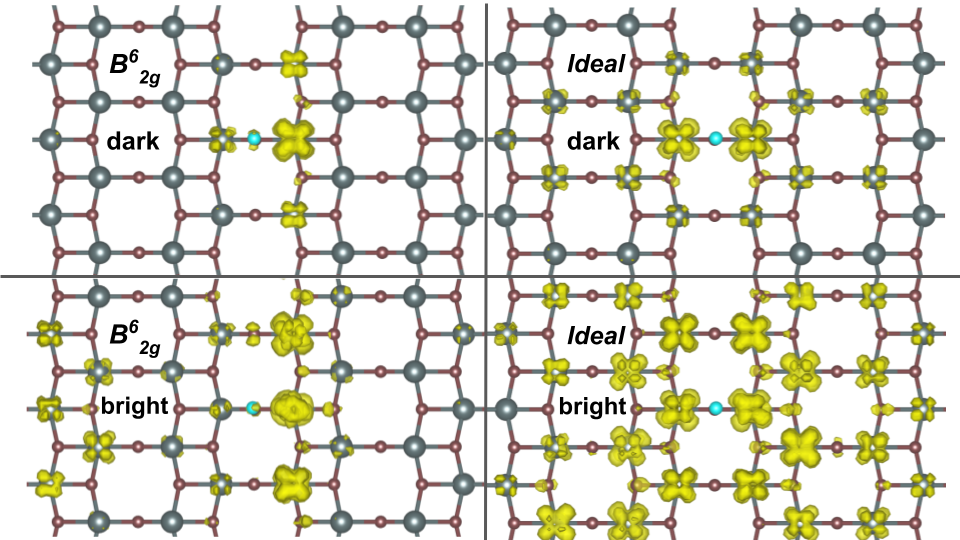}
        \end{minipage}
\end{minipage}
\caption{\small{Spatial distribution of the wavefunction $|\Psi_\lambda(\bfr_h,\bfr_e)|^2$ of the lowest energy dark (upper panels) and bright excitons in $y$ polarization (bottom panels)
for fixed position of the hole $\rvec^0_h$ close to a bridge oxygen O$_b$ atom (light blue dot): distorted crystal structure along the $B^6_{2g}$ mode  (left panels) and  ideal crystal structure (right panels).
The isosurface value corresponds to  2\% of the maximum.
}\label{fig:WXC_wf_b2g6}}
\label{fig:exciton-nature}
\end{figure*}

The distortion decreases both $E_g$ and $E^{\rm eff}_{\lambda}$ of the dark exciton by 1 eV (compare rows \textcircled{1} and \textcircled{2}), meaning that the change in binding energy $E_b$ of the dark exciton is indeed due to a change in $\langle W_{\lambda} \rangle$ [see Eq. \eqref{eq:Eb}]. 
$E^{\rm eff}_{\lambda}$ of the bright exciton, instead, decreases by as much as 1.5 eV, which cancels to a large extent the 0.7 eV decrease of the bright $\langle W_{\lambda} \rangle$, leading to a similar change in binding energy as for the dark exciton, but with a different interpretation: visibly, the nature of the bright exciton is significantly affected by the distortion.

Again, it is interesting to examine whether this change is due to a change of screening. Table \ref{tab:exc-dist} therefore compares $\langle W_{\lambda} \rangle$, see Eq. \eqref{eq:w-lambda}, calculated  by using the matrix elements
$\langle vc \kvec | W |v'c'\kvec' \rangle$ in the ideal (third column) or
the distorted (forth column) structure. 
For the exciton wavefunction  $A_{\lambda}^{vc\kvec}$, in Eq. \eqref{eq:w-lambda} we also use the result of the excitonic Hamiltonian constructed with the QP electronic structure of the distorted system and with either the screened interaction $W_{\rm dist}$ of the distorted  structure (row \textcircled{2}) or $W_{\rm ideal}$ of  
the ideal  structure (row \textcircled{3}). All the four possible combinations entering Eq. \eqref{eq:w-lambda} always give the same results, both for bright and dark excitons, meaning that 
the choice of $W$ has almost no consequences, neither in the excitonic hamiltonian nor in Eq. \eqref{eq:w-lambda}. 

We can conclude this part by noting that the change of $W(\bfr,\bfr')$ due to distortion has negligible impact on the exciton and can safely be neglected, confirming that one can use the screened Coulomb interaction calculated in the ideal system. This stability is even more pronounced than in the case of the self-energy, maybe due to the fact that the exciton probes shorter distances.
This suggests that when studying bound excitons in systems with defects, disorder and at finite temperature, the screened Coulomb interaction of the ideal structure can be used, which will increase the feasibility of such studies \cite{Garbuio2006,Dong2021}.

The change in the QP energies and wavefunctions upon distortion, instead, changes the exciton wavefunction and therefore the effective gap and the matrix element of $W$, with an effect on the exciton binding energy that is very much reduced thanks to a cancellation between the two changes. The fact that the exciton wavefunction changes under the distortion can be directly appreciated in Fig. \ref{fig:exciton-nature}: both the dark and the bright exciton have a tendency to decrease their extension in $x$ direction and increase in $y$ direction, with the distorted distribution of the electron being non-symmetric with respect to the position of the hole due to non-symmetric V-O$_b$-V bond lengths.
Examining the exciton mean radius over the supercell volume $\Omega$ for each direction $\alpha=\{x,y,z\}$  $$\langle | r^{\alpha}_{\lambda}| \rangle = \frac1{\Omega} \int_{\Omega} d \rvec_e |r^{\alpha}_e-r^{0 \alpha}_h| |\Psi_{\lambda}(\mathbf{r}^0_h,\mathbf{r}_e)|^2$$
can give a quantitative estimate of the localisation character of the exciton.  
For the ideal structure the exciton radii of the dark and bright excitons form an ellipsoid with axes: (5.3, 2.0, 2.8){\AA} and (8.8, 3.3, 4.0)\AA, respectively. In the distorted structure they become (4.7, 2.2, 3.3){\AA} and (7.8, 3.3, 3.9)\AA, respectively.
Although these effective radii show that the change is less dramatic than what the picture would maybe suggest, the effect is still sizeable and one should be aware of it and not judge on the absence of changes by just looking at the binding energy.

\section{Conclusions}
\label{sec:conclusion}

We have analysed the electron excitation spectra of {\vo} and investigated in detail the different impact of structural deformations on the optical and dielectric properties of {\vo}.
While excitons in optical absorption spectra are sensitive to the local atomic environment, especially through modifications of the band structure, plasmons in loss function spectra see mainly the global environment and are less affected by local distortions.
In particular, we have shown that the modifications in the optical spectra are due to counteracting changes. 
The impact of the structural distortion on the effect the electron-electron interaction on the band structure is the result of cancellations, to a very large extent, between the changes of the GW self-energy and of the Kohn-Sham xc potential. As a result, the main effect of the distortion is already captured at the level of the Kohn-Sham band structure.
Similarly, the change in the effective band gap largely compensates the change in the electron-hole interaction in the BSE exciton hamiltonian. As a result  the exciton binding energy turns out to be less affected, even though the nature of the exciton can be significantly impacted by the local distortion.
Finally, our results show that, and why,  for the simulation of excitons in materials with low-symmetry crystal structures, or in presence of disorder, the screened Coulomb interaction of a higher-symmetry structure can be used, promising a large computational gain.

\section*{Acknowledgements}
We thank Claudia R\"odl for  suggesting us to look into the particular convergence of the BSE JDOS and absorption spectra, and for insightful discussions.
This work benefited from the support of EDF in the framework of the research and teaching Chair ``Sustainable energies'' at Ecole Polytechnique. Computational time was granted by GENCI (Project No. 544). WRLL was supported by the US Department of Energy-Basic Energy Sciences under grant  No. DE-SC0008933. 

\appendix
\section{Additional computational information}

It is useful to 
look closer to two aspects of the calculations in order to combine efficiency and accuracy.

\subsection{$\bfk$-point convergence}
 \label{app:kpoint} 
 
 \begin{figure}[t]
\center
\begin{minipage}[b]{\columnwidth}
\center
        \begin{minipage}[b]{\columnwidth}
        \includegraphics[width=0.7\columnwidth]{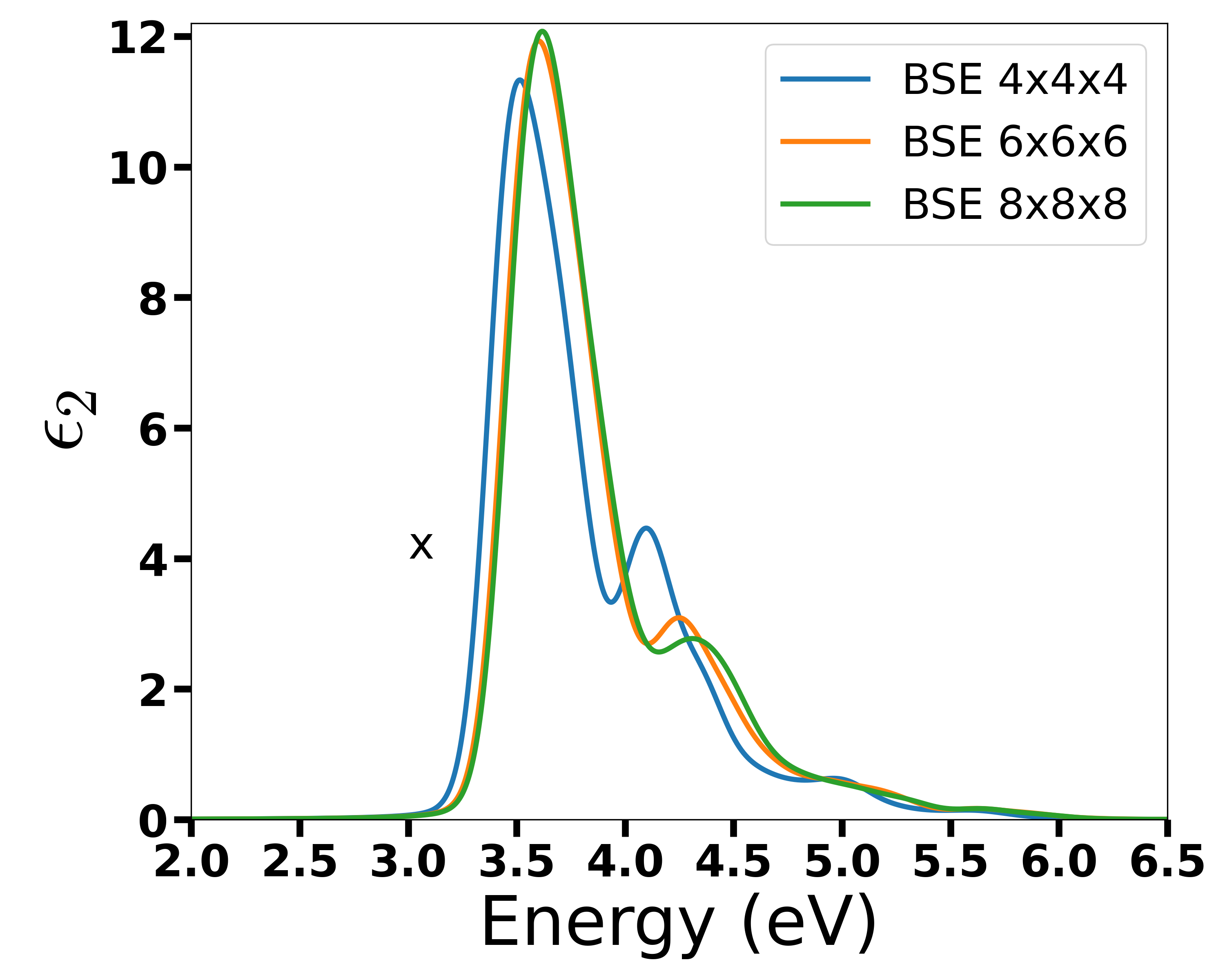}
        \includegraphics[width=0.7\columnwidth]{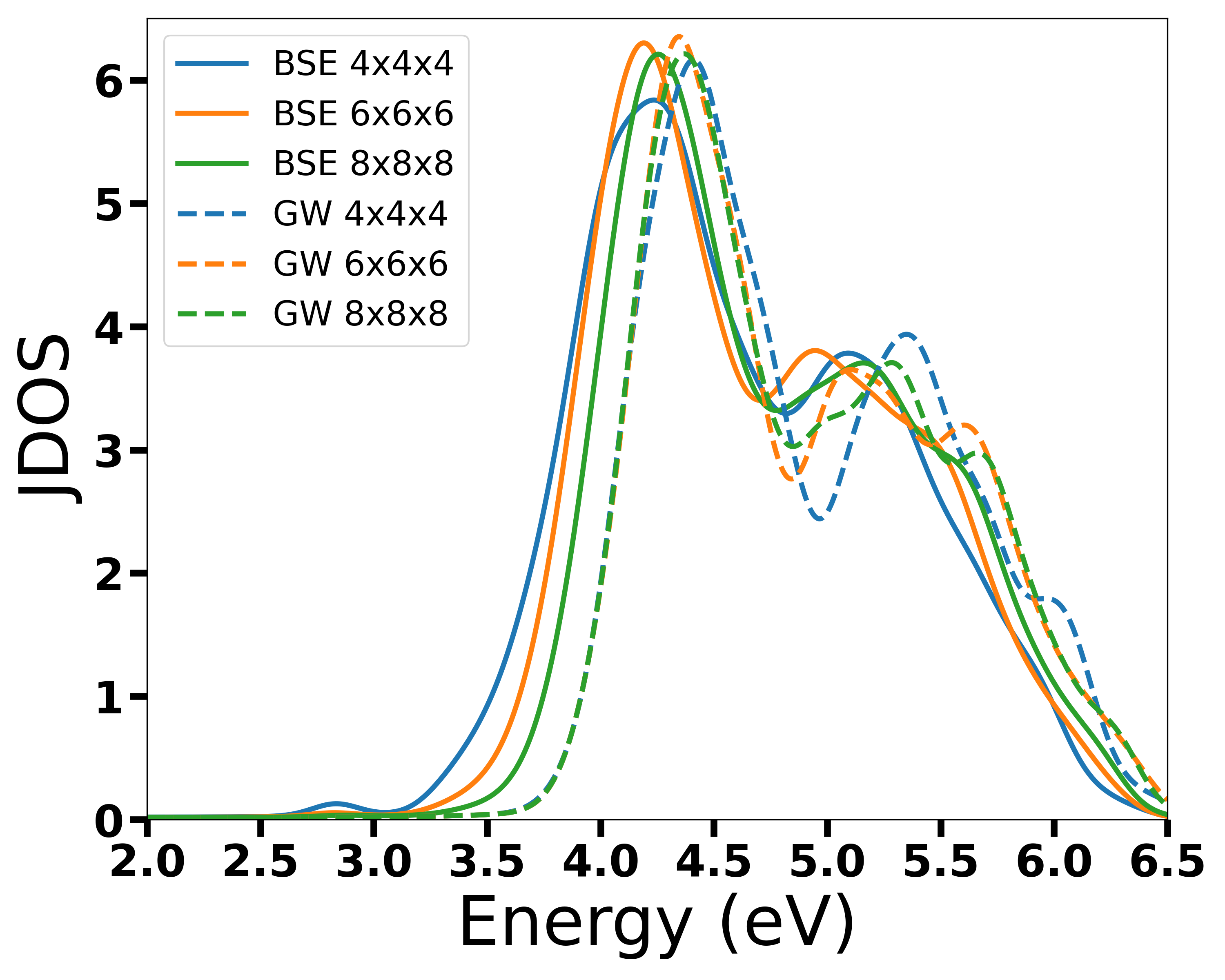}
        \includegraphics[width=0.7\columnwidth]{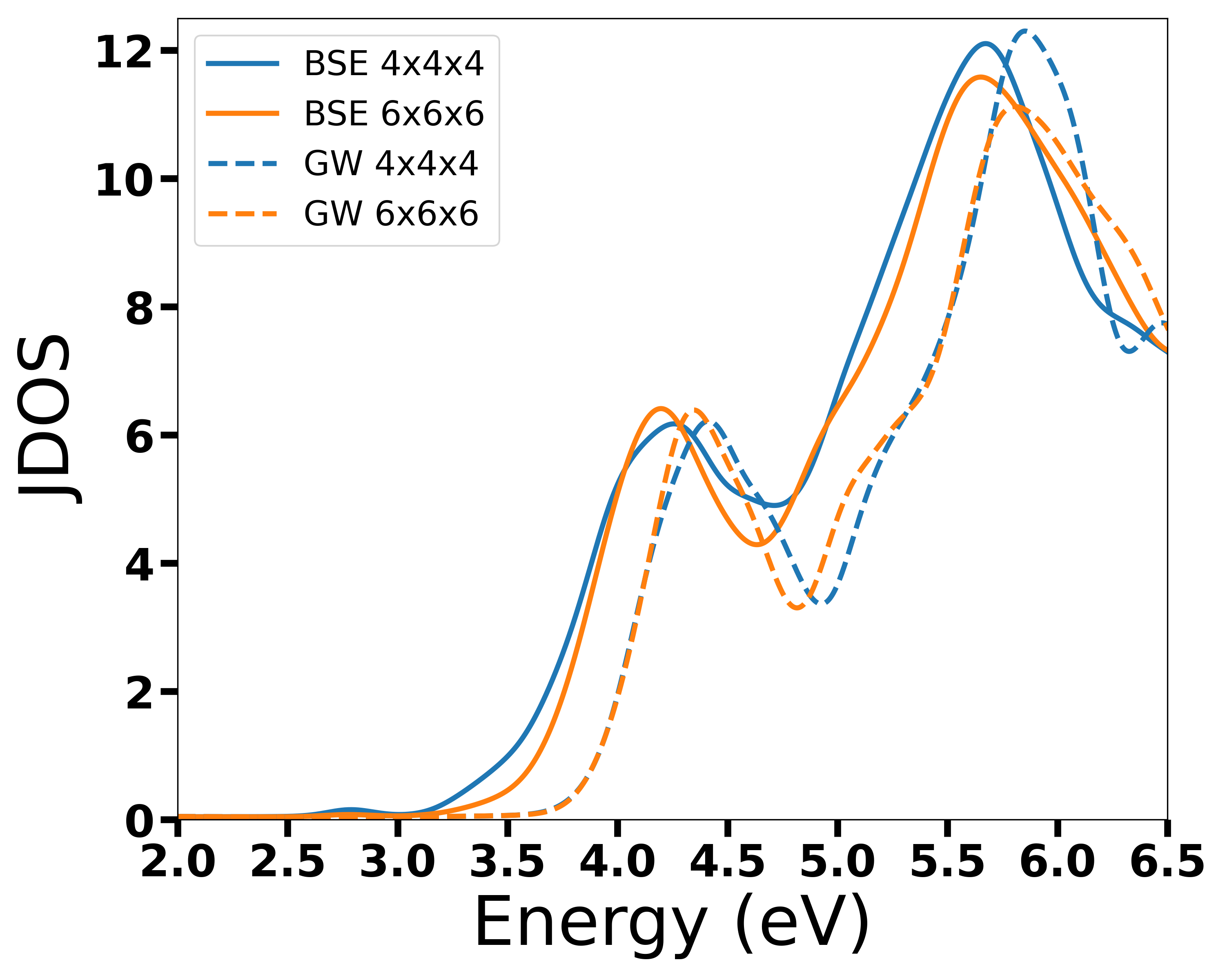}
        \end{minipage}
\end{minipage}
\caption{\small{Convergence with respect to the Brillouin zone sampling of spectral features of {\vo} in the ideal crystal structure. Upper panel: BSE absorption spectra along $x$ polarisation direction. Middle and lower panel:  JDOS of BSE and GW-IPA for different grids. Middle panel, 7 occupied bands and 4 unoccupied bands. Lower panel: 13 occupied and 10 unoccupied bands.  }}
\label{fig:JDOS}
\end{figure}

It is important to note that the precision required depends very much on the feature of interest.  
The convergence with the $\bfk$-point grid is a particularly good example for this point, and since it has not been treated exhaustively in Ref.\cite{Gorelov2022}, we will give details here.
The upper panel of Fig. \ref{fig:JDOS} shows the $\bfk$-point grid convergence of the absorption spectrum in the BSE, here calculated using only 7 occupied and 4 unoccupied bands. Clearly, the main peak is unchanged whether a $6\times6\times6$ or a $8\times8\times8$ grid is used, and it is already very well described using  a $4\times4\times4$ grid. This quite coarse grid yields hence absorption spectra that can be discussed safely. Instead, using the same grid is more delicate when it comes to details of the JDOS, which is not a measurable quantity, but which is sometimes used for discussion, such as in Ref.\cite{Gorelov2022}. Therefore, we show in the present article the convergence of the JDOS in Fig. \ref{fig:JDOS}. While, as in the case of the spectrum, the main peaks of the JDOS are well identified by the coarsest $\bfk$-point grid, the slope on the low-energy side (< $\sim$ 3.7 eV) of the main peak is changing slowly but noticeably with the grid size, in particular at the onset. This finding, which is closely related to the convergence difficulties pointed out in \cite{Fuchs2008}, is independent of the number of bands used, as a comparison of middle and lower panels shows. This delicate behavior of the onset is true only in the BSE JDOS: the analogous changes in the GW-IPA JDOS can hardly be seen. Such a convergence problem in the BSE may be explained by a situation where states $\lambda$ have coefficients $A_{\lambda}^{vc{\bfk}}$ that are a very narrow distribution with ${\bfk}$, narrower than the difference between ${\bfk}$-points of the grid. When one uses a discrete ${\bf k}$-point grid, $W(\bfq)$ is supposed to be constant with $\bfq$ over each grid interval or, as it is routinely done for the point $\bfq=0$, one replaces $W(\bfq)$ by the average, i.e., an integral over the interval.  This leads to an overestimation of the contribution of the long-range part of the screened Coulomb interaction \cite{Fuchs2008}. The error is not noticeable in simple semiconductors, but can play a role in materials such as the present one, where the electron-hole interaction is very strong. Narrow distribution of coefficients occur in the electron-hole continuum, whereas bound excitons have a broader $\bfk$-distribution (see, e.g., Fig. \ref{fig:BS_dist}). When the distribution is broader than the intervals of the $\bfk$-grid, the piecewise treatment of $W(\bfq)$ is justified.
Since the intensity of the bright excitonic states is much larger than that of unbound states, this explains why the spectrum converges quickly. 
Moreover, contrary to the spectrum, in the JDOS the weight of discrete states should tend to zero with increasing number of $\bfk$-points.
One can therefore safely discuss the absorption spectrum with a $4\times 4\times 4$-grid, but not the entire JDOS, especially in the region of the onset between 2.5 and $\sim$ 3.7 eV. Indeed, the presence of a number of bound excitons over a large spectral range, whose energy is stable with the $\bfk$-point sampling, is confirmed even by the larger grids, which explains the stability of the spectrum and, despite the convergence issue, confirms the conclusions of Ref.\cite{Gorelov2022} concerning the difficulty to determine the onset of the continuum.
 
\subsection{Scissor-correction approximation}
\label{app:scissor}

\begin{figure*}[ht]
\center
\begin{minipage}[b]{2.\columnwidth}
\center
        \begin{minipage}[b]{\columnwidth}
        \includegraphics[width=0.9\columnwidth]{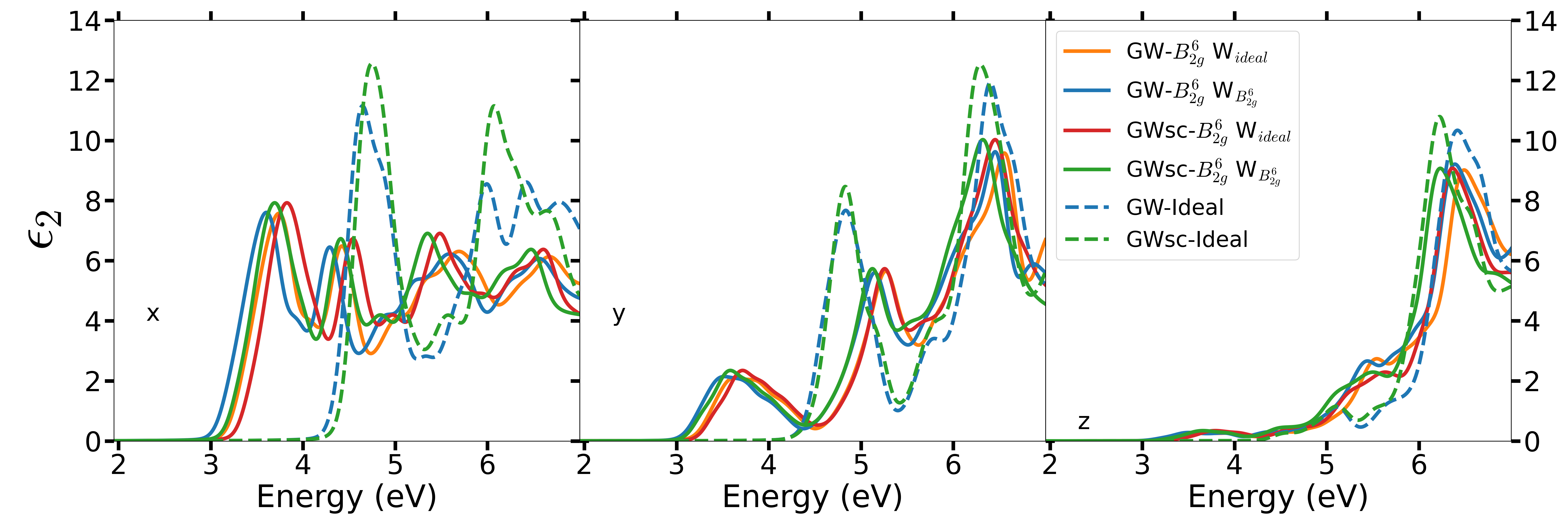}
        \end{minipage}
\end{minipage}
\caption{GW-IPA absorption spectra, in the 3 cartesian polarization directions, (left) $x$, (middle) $y$ and (right) $z$, of the crystal structure distorted along the $B_{2g}^6$ mode with the largest displacements (solid lines)
  compared to those of ideal crystal structure (dashed lines): full GW-IPA (blue lines) and scissor-corrected (sc) GW-IPA (green lines).
For the distorted structure, the full GW spectra (orange lines) and scissor-corrected (sc) GW spectra (purple lines) are also obtained with the screened Coulomb interaction $W_{\rm ideal}$ calculated in the ideal {\vo}.
\label{fig:ABS_B2g6_2}}
\end{figure*}

The spectra in Fig. \ref{fig:ABS_B2g6_2}  are calculated within the GW-IPA, using different ingredients, for the ideal {\vo} (dashed lines) and the structure with the largest distortion for the $B_{2g}^6$ mode (solid lines).
In the full calculation (blue curves) the spectra are obtained taking into account the GW corrections (see Eq. \eqref{eq:GW}) for all the bands and $\bfk$ points included in the response function.
Alternatively, the GW corrections have been approximated in the spectra (green curves) by a rigid scissor correction for the conduction states, equal to the GW band gap opening at the $\Gamma$ point. 
We find that the effect of the scissor correction approximation with respect to the full calculation is small. In both cases we find QP corrections with a spread of 0.3-0.4 eV over the different states, but without a clear tendency, such that a large part of the effect averages out in the spectrum. Since the main peaks in the absorption spectrum are less sharp in the distorted than in the ideal structure, this averaging out of the difference with respect to a simple scissor correction is even more efficient.
Most importantly, however, the error due to the scissor is in all cases much smaller than the difference between the spectra for the distorted and ideal structures.

In Fig. \ref{fig:ABS_B2g6_2} we also verify, for the  GW-IPA spectra of the distorted structure, the use of the screened Coulomb interaction $W_{\rm ideal}$ of the ideal structure instead of $W_{\rm dist}$ of the distorted structure. The comparison is excellent for both the  spectra with the full GW corrections (compare blue and orange solid lines) and the spectra obtained with the corresponding scissor corrections (compare green and purple solid lines).

\bibliography{main}

\end{document}